\documentclass[fleqn,usenatbib]{mnras}

\usepackage{newtxtext,newtxmath}
\usepackage[T1]{fontenc}
\usepackage{ae,aecompl}
\usepackage{soul}
\usepackage[normalem]{ulem}

\usepackage{float}

\usepackage{graphicx}	% Including figure files
\usepackage{amsmath}	% Advanced maths commands
\usepackage[usenames,dvipsnames]{xcolor}

\hypersetup{draft}

\newcommand{\Mpc}{{\textrm{ Mpc}}}
\newcommand{\hMpc}{{\textrm{ $h^{-1}$Mpc}}}
\newcommand{\kms}{{\textrm{ km\,s$^{-1}$}}}

\newcommand{\hMsun}{{\textrm{ $h^{-1}$M$_{\sun}$}}}
\newcommand{\Jdir}{\ensuremath{\hat{\mathbf{J}}}}
\newcommand{\ea}{\ensuremath{\hat{\mathbf{e}}_1}}
\newcommand{\eb}{\ensuremath{\hat{\mathbf{e}}_2}}
\newcommand{\ec}{\ensuremath{\hat{\mathbf{e}}_3}}
\newcommand{\ei}{\ensuremath{\hat{\mathbf{e}}_i}}
\newcommand{\ta}{\ensuremath{\hat{\mathbf{t}}_1}}
\newcommand{\tb}{\ensuremath{\hat{\mathbf{t}}_2}}
\newcommand{\tc}{\ensuremath{\hat{\mathbf{t}}_3}}
\newcommand{\ti}{\ensuremath{\hat{\mathbf{t}}_i}}
\newcommand{\ia}{\ensuremath{\hat{\mathbf{i}}_1}}
\newcommand{\ib}{\ensuremath{\hat{\mathbf{i}}_2}}
\newcommand{\ic}{\ensuremath{\hat{\mathbf{i}}_3}}
\newcommand{\ii}{\ensuremath{\hat{\mathbf{i}}_i}}
\newcommand{\Wh}{\textrm{H}}
\newcommand{\Lh}{\textrm{L}}
\newcommand{\Mh}{\textrm{M}}

\newcommand{\nexus}{\textsc{nexus+}}

\definecolor{ao}{rgb}{0.0, 0.5, 0.0}

\title[Evolution of the halo spin--filament alignment]{Deviations from tidal torque theory: evolution of the halo spin--filament alignment}

 \author[L\'opez et al.]{Pablo L\'opez,$^{1,2}$\thanks{E-mail: plopez@unc.edu.ar}
 Marius Cautun,$^{3}$
 Dante Paz, $^{1,2}$
 Manuel Merch\'an, $^{1,2}$
 \newauthor{and
 Rien van de Weygaert$^{4}$}
\\
$^{1}$Observatorio Astron\'omico de C\'ordoba, Universidad Nacional de C\'ordoba (UNC), Francisco N. Laprida 854, C\'ordoba, Argentina\\
$^{2}$Instituto de Astronom\'ia Te\'orica y Experimental, CONICET-UNC, Laprida 922, C\'ordoba, Argentina\\
$^{3}$Leiden Observatory, Leiden University, PO Box 9513, NL-2300 RA Leiden, the Netherlands\\
$^{4}$Kapteyn Astronomical Institute, University of Groningen, PO Box 800, 9747 AD, Groningen, The Netherlands\\
}

\date{Accepted XXX. Received YYY; in original form ZZZ}

\pubyear{2020}

\begin{document}
\label{firstpage}
\pagerange{\pageref{firstpage}--\pageref{lastpage}}
\maketitle

\begin{abstract}
    The alignment between halo spins and the cosmic web is still poorly understood despite being a widely studied topic. Here, we study this alignment within the context of tidal torque theory (TTT) and deviations from it. To this end, we analyze the evolution of the shape and spin direction of proto-haloes, i.e. of all the volume elements associated to a $z=0$ halo, with respect to the present-day filaments. We find that the major axis of proto-haloes undergoes a major change, from being strongly perpendicular to the filament spine in the initial conditions, to being preferentially aligned at the present time. In comparison, the spin orientation shows only a mild evolution: it starts slightly parallel to the filament spine, but the subsequent evolution, up to $z{\sim}1$, gradually changes its orientation to preferentially perpendicular. In order to analyze these signals in the TTT framework, we split the haloes according to their net spin growth with respect to the median TTT expectation, finding a clear correlation with the spin--filament alignment. 
    At the present time, haloes whose spin grew the most are the ones most perpendicular to the filament spine, while haloes whose spin grew below the median TTT expectation are typically more aligned. 
    The dependence of spin directions on net spin growth is already present in the initial conditions, and gets further modified by late-time, $z<2$, evolution. Also, spin directions mildly deviate from the TTT predictions even at high redshift, indicating the need for extensions to the model.
\end{abstract}


%%%%%%%%%%%%%%%%%%%%%%%%%%%%%%%%%%%%%%%%%%%%%%%%%%

%%%%%%%%%%%%%%%%% BODY OF PAPER %%%%%%%%%%%%%%%%%%

\section{Introduction}
\label{introduction}

For more than $70$ years, since the pioneer work of \citet{hoyle1949}, the problem of how galaxies and dark matter (DM) haloes acquired their angular momentum (AM) has constituted one of the most intriguing and addressed topics in astronomy. This question is relevant for understanding galaxy formation processes, such as the formation of gas and stellar discs, and how these processes are connected to the large-scale structure (LSS) of our Universe. Moreover, the growth and direction of the haloes' spin is directly related to the intrinsic alignment of galaxies, which affect weak lensing measurements \citep{troxelyishak2014,hikageetal2019,fabbianetal2019,copelandetal2020}, and is a source of systematic uncertainties for methods that use galaxy clustering to model the galaxy-halo relationship, such as the halo occupation distribution model \citep{zentneretal2014,rodriguezetal2015,mcewenetal2018}. Additionally, the statistical alignment of galaxy spins with the LSS has been proposed as a novel cosmological probe, such as measuring the total neutrino mass \citep{leeetal2020}. Hence, understanding the angular momentum growth of haloes and galaxies is important for a wide range of problems, such as constraining cosmological parameters and improving galaxy formation models.

One of the key ingredients for understanding the emergence of angular momentum in haloes and galaxies lies in the fact that these objects are not uniformly distributed in the Universe, but rather they compose a multi-scale, anisotropic arrangement usually referred to as the \emph{cosmic web}. This is a complex structure of densely populated nodes, linked together through elongated filaments of intermediate density, which are themselves connected by flattened walls; in addition, amidst these systems there are vast regions nearly devoid of matter \citep[][see \citealt{libeskind2018} for a recent review]{bondetal1996,weygaert&bond2008,aragoncalvoetal2010,cautunetal2014}. Visually, the cosmic web constitutes the most noticeable feature of the galaxy distribution at large-scales, as noted in early studies \citep{joeveeretal1978,delapparentetal1986,gelleryhuchra1989,shectmanetal1996} and confirmed in more recent large galaxy surveys \citep{collessetal2003,tegmarketal2004,weygaertyschaap2009,huchraetal2012,guzzoetal2014}, and this pattern is even more clearly seen in cosmological simulations \citep{springeletal2005,aragoncalvoetal2010,cautunetal2014,vogelsbergeretal2014,schayeetal2015}. These studies have shown that the cosmic web is, in fact, a manifestation of the anisotropic gravitational collapse of the tiny fluctuations present in the primordial density field, and that its growth is shaped by the large-scale tidal field \citep{bondetal1996,weygaert&bond2008}. Hence, the tidal field can be thought to be the key driver to connect phenomena on very different scales, because it is, at the same time, a consequence of fluctuations in the matter distribution and a source for that matter to acquire angular momentum as it collapses into haloes and galaxies.

This idea has been the basis of a wide and quite heterogeneous family of models and predictions on how haloes and galaxies acquire their spin, which can be summarized as part of the \emph{tidal torque theory} (TTT) approach. The hypothesis that galactic rotation arises as a consequence of the tidal interaction between neighboring disturbances was first suggested by \citet{hoyle1949} and then further developed by \citet{peebles1969}, who applied linear theory to estimate that the angular momentum acquired by an spherical proto-galaxy grows at second order. Later, \citet{doroshkevich1970} refined the idea and found that the angular momentum should actually grow at first order, due to the proto-galaxies' deviation from spherical symmetry. However, it was not until several years later that \citet{white1984} showed the calculations in detail and described the physical mechanism behind the effect: angular momentum is due to the differential alignment between the inertia tensor of a proto-galaxy and the local gravitational tidal tensor in which it is embedded. Recently, \citet{neyrincketal2019} have proposed another framework in which halo spins arise due to the residual motion of different regions inside the proto-halo, which leads to similar predictions as TTT but circumvents some of the problems of the latter.

TTT and its variants have been widely applied to model the growth of galaxies and DM haloes, such as to investigate the dynamical evolution of density peaks at high redshift \citep{fallyefstathiou1980,hoffman1986,heavensypeacock1988,steinmetzybartelmann1995,catelanytheuns1996}, the formation and orientation of galaxies \citep{lake1983,ryden1988,quinnybinney1992}, the intrinsic properties of haloes \citep{warrenetal1992} and their relation with the surrounding structure \citep{barnesyefstathiou1987}, as well as the correlations between these properties and the large-scale tidal field (see \citealt{schafer2009} or \citealt{jonesyvandeweygaert2009} for a detailed review). 

A frequently addressed topic has been the origin of spin orientations and their alignment with the LSS. For instance, it has been shown that, in the ideal situation in which the inertia tensors of proto-haloes are completely independent of their surrounding tidal field \citep{catelanytheuns1996}, the TTT mechanism would produce a preferential alignment between the spin and the intermediate axis of the local shear tensor \citep{leeypen2001}. 
When the strong correlation between the inertia tensor and the local tidal field has been taken into account in numerical simulations, this alignment has been found to be weaker, but nonetheless still present. \citep{leeypen2000,porcianietal2002b}.
However, numerical simulations have also shown that spin directions at present time can differ by tens of degrees from their orientation at the linear stages of structure formation (e.g. the initial conditions of numerical simulations), which is when the basics assumptions of TTT are valid \citep{porcianietal2002a}. Hence, it is not clear to what extent the predictions from the model can be used when analyzing present day alignments.

In fact, a recurring conclusion among TTT-based analysis is that the model adequately describes the mean or the median evolution of a DM halo population, but that it is not capable of accurately predicting the angular momentum of individual haloes. As already pointed by \citet{white1984}, the two main causes for this limitation are that the final spin depends critically on how and when the TTT process is terminated and that non-linear effects cause the evolution of angular momentum to vary considerably from case to case. In other words, TTT is adequate to predict the evolution of angular momentum only during the linear and quasi-linear stages of structure formation \citep{sugermanetal2000,porcianietal2002a,porcianietal2002b}, but its effects tend to be erased at late stages, when haloes and galaxies are greatly affected by non-linear processes such as mergers \citep{vitvitskaetal2002,bettyfrenk2016} and the emergence of vortical flow fields \citep{libeskindetal2012}, although it has been shown that the latter can be reconciled with the TTT if certain assumptions are made in order to account for the anisotropy of the larger scale environments \citep{codisetal2015,laigleetal2015}.

Altogether, studies indicate that the alignments of halo spin with their surrounding structure are not as simple as predicted by linear theories such as TTT. Several authors have studied this relation in numerical simulations \citep[e.g.][]{bailinysteinmetz2005,aragoncalvoetal2007,hahnetal2007,sousbieetal2008,codisetal2012,libeskindetal2013,foreroromeroetal2014,wangykang2017,veenaetal2019,lopezetal2019,pereyraetal2020} and observational catalogs \citep[e.g.][]{leeyerdogdu2007,pazetal2008,jonesetal2010,tempelylibeskind2013,zhangetal2015,welkeretal2019,bluebirdetal2020}. One of the most reproduced results refers to the mass dependency of the haloes' spin--filament alignment. It can be summarized in the finding that less massive haloes tend to have their spins aligned with the axis of their host filament, whereas higher mass haloes have their spin preferentially perpendicular to the filament axis. The passage from one regime to the other occurs at a particular \emph{spin transition} mass, which has been reported to be $\sim 10^{12}\hMsun$, with a significant dependence of this value on the methods used to identify the cosmic web environments and on the properties of these environments, such as thickness, length and local density \citep[e.g.][]{hahnetal2007,aragoncalvoetal2007,codisetal2012,aragon2014,veenaetal2018,veenaetal2020}.

\subsection{Scope of this work}
\label{sec:scope}
Many authors have studied the evolution of the alignment between halo spins and the cosmic web with the goal of understanding the major processes driving this correlation \citep[e.g.][]{hahnetal2007,aragoncalvoetal2007,codisetal2012,wangykang2017,Wang2018a,Wang2018b,veenaetal2020}. The most common approach involves identifying the virialized haloes and the cosmic web at each redshift and following how the halo spin--web alignment changes in time. However, there are two aspects of this methodology that makes it challenging to interpret within the TTT framework. First, the cosmic web changes in time \citep[e.g.][]{cautunetal2014} and that, in turn, can lead to changes in the spin--web alignment \citep{wangykang2017}. Secondly, the angular momentum growth of virialized halo progenitors is not easily interpreted in the TTT approach. TTT applies to proto-haloes (i.e. all the fluid elements associated to the present-day halo) and, at any given redshift, a proto-halo includes the main halo progenitor as well as additional lower mass haloes and diffuse material that are yet to be accreted into the main progenitor \citep[e.g. see][for examples of halo merger trees]{Lacey1993}. According to this definition, the mass of proto-haloes is constant in time, and thus their angular momentum grows only due to external torques. In contrast, the spin growth along the main branch is a combination between the torques acting on the main progenitor and the angular momentum brought by newly accreted mass, which can be very different \citep{Zavala2016,Liao2017}.

In this paper, we study the alignment between DM halo spins and the cosmic web orientation from the perspective of TTT and deviations from it. In particular, we aim to investigate to what extent the present-day halo spin--filament alignment is already set in the initial condition and how this alignment is affected by the subsequent angular momentum growth that is not captured by TTT. This work is different from previous studies in two major aspects that, as we argued in the previous paragraph, are essential for interpreting the results within the TTT framework. Firstly, we follow the spin growth of proto-haloes, that is the set of all DM particles that are associated to a present-day halo.  
Secondly, we define the cosmic web only at $z=0$, and correlate the proto-halo spins only against the $z=0$ web. This has the advantage of drastically simplifying the problem since now any redshift change in spin--filament alignment is due to changes only in the proto-halo spin.

More concretely, our paper addresses the following three questions:
\begin{itemize}
    \item What is the TTT prediction for the present-day alignment between halo spins and their host filaments?
    \item How is this alignment affected by the angular momentum growth not captured by TTT?
    \item And, within the same context, is there a correlation between the halo angular momentum growth and its alignment with the host filament?
\end{itemize}
To answer these questions, we study Friends-of-Friends (FOF) haloes in a large volume and high resolution DM-only simulation. We trace back in time the particles associated to each $z=0$ FOF halo and correlate their spin at each redshift with the $z=0$ filaments identified by the \textsc{nexus+} web finder \citep{cautunetal2013}. We limit our analysis to filaments since they contain the majority of haloes and are easily identifiable in both cosmological simulations and observations \citep[e.g.][]{cautunetal2014,libeskind2018}.

In the second part, we study if the spin--filament alignment is related to the amount of spin growth. For this, we employ the \citet{lopezetal2019} approach in which haloes are split into categories depending on the amount of spin growth they experienced with respect to the median TTT prediction for their halo mass. This was motivated by the \citeauthor{lopezetal2019} results that found clear correlations between spin growth, halo clustering and LSS. The haloes which experienced the largest spin growth typically formed late, are more clustered, and have their spin preferentially perpendicular to the filament axis. In contrast, haloes that experienced the least amount of spin growth formed early, are less clustered, and have their spin preponderantly along the filament axis. 

The paper is organized as follows: in \autoref{sec:TTT_outline} we outline TTT and present the expectations for both the halo AM growth and direction; in \autoref{sec:methods} we describe the numerical simulation, the halo properties and the method we use to identify the cosmic filaments; in \autoref{sec:alignment_all} we show the time evolution of the spin--filament alignment for \emph{all} haloes and, specifically, the discrepancy between the TTT predictions and the present-day measurements; in \autoref{sec:alignment_growth} we present the main results of our work: we show that the AM growth of haloes is strongly correlated with the present-day spin--filament alignment and its time evolution; finally, in \autoref{conclusions}, we summarize our main results, explore possible causes for the observed trends and briefly discuss implications for both TTT-based and non-linear analysis. 

\section{Brief outline of TTT}
\label{sec:TTT_outline}

TTT predicts the growth of angular momentum in the linear regime of structure formation, when proto-haloes corresponds to small density fluctuations that evolve according to the Zel'dovich formalism \citep{zeldovich1970}. Let us consider a proto-halo whose moment of inertia with respect to its center of mass is denoted by $\mathbfss{I}$ and that is located in a tidal field, $\mathbfss{T}$. According to TTT, the angular momentum of this proto-halo can be expressed at time, $t$, as \citep{white1984}:
\begin{equation}
    \label{eq:TTT}
    J_i(t)=a^2(t)\dot{D}(t) \ \epsilon_{ijk}T_{jl}I_{lk},
\end{equation}
where $a(t)$ is the scale factor, $D(t)$ is the linear growth factor that describes how density perturbations evolve in time, the dot denotes a derivative with respect to cosmic time $t$, and $\epsilon_{ijk}$ represents the fully antisymmetric rank-three tensor (for a more detailed description of the TTT, see e.g. section 2 in \citeauthor{lopezetal2019} \citeyear{lopezetal2019} and references therein). 

Equation \ref{eq:TTT} states that during the early stages of structure formation the AM is the tensor product between the proto-halo inertia tensor and the surrounding tidal field in the initial conditions. Within the TTT approach only the magnitude of the AM grows, whereas the spin direction is the same at all times. The time dependence for the AM growth is determined solely by the factor $a^2(t)\dot{D}(t)$, which in an Einstein-de Sitter (EdS) universe is $\propto t$. 
The TTT prediction for spin growth, i.e. Equation \ref{eq:TTT}, is valid as long the following four assumptions hold \citep{porcianietal2002a}: i) the flow is laminar, i.e. no shell-crossing has taken place, ii) the velocities are well approximated by the Zel'dovich formalism, iii) the gravitational potential at each point of the proto-halo is well approximated by the second order Taylor expansion with respect to the halo centre, and iv) non-linear processes do not lead to a large effect on the total AM. During the late-time evolution of the proto-halo, many of these assumption will not be valid and thus, at low redshift, TTT will fail to properly predict the proto-halo spin. 

The halo AM is expected to grow until the turn-around time, after which the halo shrinks and the tidal torques acting on it become progressively less efficient \citep[][although close interactions, such as another halo fly-by, can still lead to a large AM change -- \citealt{bettetal2007,bettyfrenk2016}]{peebles1969}. In the case of TTT, \citet{porcianietal2002a} have shown that the framework gives an unbiased prediction of the halo spin magnitude if the AM growth stops at ${\sim}60\%$ of the turn-around time. However, while TTT provides a reasonable description for the mean spin magnitude of the halo population, it does more poorly when predicting the spins of individual haloes, whose values are dominated by non-linear processes \citep[e.g.][]{porcianietal2002a,lopezetal2019}.

\begin{figure*}
	\includegraphics[width=2\columnwidth]{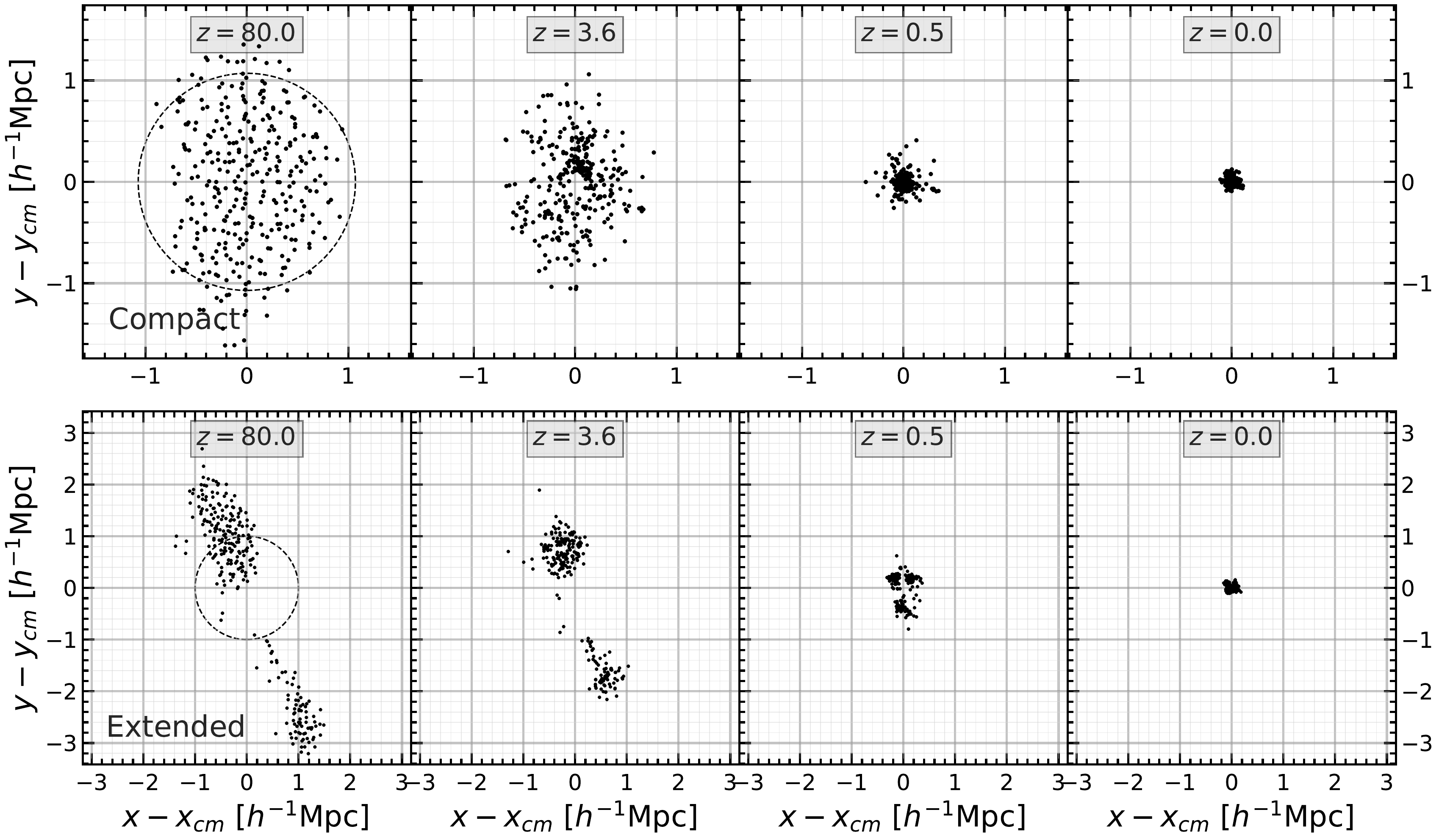}
	\vskip -.3cm
    \caption{Illustration of the proto-halo evolution for two objects with a mass of ${\sim}3.5\times10^{12}\hMsun$. The proto-haloes consists of all DM particles associated to the present-day haloes. The figure shows two such objects at redshifts 80, 3.6, 0.5 and 0, from left to right, and the particle distributions are centered at the position of their centre of mass at each time. The majority of proto-haloes are rather compact, similar to the one shown in the top row, however a small fraction can be highly extended, such as the case depicted in the bottom row. The dashed circles at $z=80$ illustrate the radius of the sphere needed to enclose the same mass as the object. Notice that the axes in the two rows have different ranges.
    \label{fig:halo_2xn}}
\end{figure*}

In this work, we are mainly studying the orientation of a proto-halo's AM. In this respect, TTT predicts that the spins show a modest degree of alignment with the initial tidal tensor: with the spins being oriented preferentially along the second and third eigenvectors of the tidal field \citep{porcianietal2002b}. However, non-linear effects erase most of this correlation, and the $z=0$ spins shows a much weaker alignment with the initial tidal field. This is a consequence of the late-time processes that can have a large impact on the spin direction: the average misalignment angle between the $z=0$ spin orientation and the TTT prediction is ${\sim}50^\circ$ \citep{porcianietal2002a}.

However, the spin changes due to non-linear processes are likely to be correlated to the non-linear cosmic web. For example, the late-time tidal field is strongly correlated to the anisotropies of the mass distribution, i.e. the cosmic web. Similarly, more dramatic processes, such as fly-by encounters, are more likely to take place along the filament axis in which a halo is embedded \citep{vanhaarlem&vandeweygaert1993}. These aspects motivate one of the central questions that we will study here: To what extent is the late change in spin orientations correlated to the late-time cosmic web?

%-----------------------------------------------------------------------------------

\section{Methods}
\label{sec:methods}

\subsection{Numerical simulation and halo catalogue}
\label{sec:methods:simulation}
Our analysis makes use of a dark matter-only simulation in a periodic box of $400\hMpc$ side-length that has been run using the \textsc{gadget 2} code developed by \citet{springelgadget2005}. The simulation models the growth of cosmic structures using
$1600^{3}$ dark matter (DM) particles, each with a mass, $m_\mathrm{p}=1.18219\times10^{9}\hMsun$. The initial conditions were generated at $z=80$ and the particle distribution was evolved down to present time, $z=0$. To follow the growth of structure, we saved 205 outputs. The simulation used the \citet{plankcollaboration2018} cosmological parameters, which consist of a matter density $\Omega_\mathrm{m} = 1 - \Omega_{\Lambda} = 0.315$, where $\Omega_{\Lambda}$ is the density parameter of dark energy, Hubble constant $H_0 = 67.4 \kms\Mpc^{-1}$, and normalization parameter $\sigma_{8} = 0.811$.

We have identified $\sim 6.8\times10^{6}$ DM haloes at the last snapshot of the simulation using the standard Friends-of-Friends \citep[FOF;][]{Davis1985} algorithm, with a percolation length of 0.17 times the mean particle separation. We characterize the haloes by their mass, which is given by the total number of particles times the particle mass. From this population, we have selected those haloes whose mass is within the range $3\times10^{11}$ to $3\times10^{13}\hMsun$. The result is a sample of $\sim 6.5\times 10^{5}$ haloes with a minimum number of $250$ particles, an amount sufficient to avoid biases in the determination of dynamic properties that usually arise when working with too few particles \citep[e.g. see][]{pazetal2006,bettetal2007}. 

As we argued in the introduction and \autoref{sec:TTT_outline}, we are interested in following the evolution of proto-haloes, which consist of all the DM particles associated to the present-day FOF halo. In order to follow the proto-haloes, we have used the particle IDs, which consist of an unique numerical label assigned to each particle in the simulation. For each $z=0$ halo, we have identified all the particles associated to it and then traced them back to earlier times to obtain the proto-halo formation history. \autoref{fig:halo_2xn} illustrates two examples of proto-haloes, showing the distribution of particles in the initial condition (left-most panel), two intermediate redshifts, and at present-day. Most proto-halo Lagrangian patches are rather compact, as highlighted in the top-row of the figure. However, a small fraction of haloes are formed from the collapse of extended and elongated patches, as illustrated in the bottom row of the figure.
We have defined \emph{extended} proto-halos as those for which more than half of their DM particles are found outside their Lagrangian radius (i.e. the radius of a sphere enclosing a mass equal to the one of the proto-halo). The fraction of such objects is very low, less than $2\%$ for haloes more massive than $10^{13}\hMsun$, and increases to ${\sim}6\%$ for the lowest mass objects considered in this work. Extended proto-haloes end up forming, at the present time, haloes with minor to major axis ratios that are typically ${\sim}40\%$ lower than their \emph{compact} counterparts.

\subsection{Proto-halo shape and spin}
\label{sec:methods:shape_and_spin}
In this paper we study two proto-halo properties, shape and spin, and how they evolve in time. The proto-halo shape at time $t$ is given by the moment of inertia of the particle distribution, that is
\begin{equation}
    \label{eq:IT}
    I_{ij} = %\frac{m_p}{N_\mathrm{h}}
    m_p \sum_{\alpha = 1}^{N_\mathrm{h}}
    x_{\alpha; \ i}(t) \ x_{\alpha; \ j}(t),
\end{equation}
where $N_\mathrm{h}$ is the number of particles in the halo and $x_{\alpha; \ i}(t)$ represents the $i$-th component of the displacement vector of the $\alpha$-th particle with respect to the centre of mass at time $t$. Note that at each redshift the summation is over the same set of particles and the only variables that change are the particle coordinates. We then proceed to find the principal axes of the proto-halo, which have been obtained by diagonalizing the moment of inertia. The eigenvectors of $\mathbfss{I}$, which we denote with \ia{}, \ib{} and \ic{}, correspond to the proto-halo major, intermediate, and minor axes, respectively. The principal axes will be used to study the alignment between proto-halo shape and the present-day cosmic web.

The proto-halo angular momentum is calculated as
\begin{equation}
    \label{eq:spin}
    \mathbfss{J} =  %\frac{m_p}{N_\mathrm{h}}
    m_p \sum_{\alpha = 1}^{N_\mathrm{h}}
    \mathbfss{x}_{\alpha}(t) \times \mathbfss{v}_{\alpha}(t),
\end{equation}
where $\mathbfss{x}_{\alpha}(t)$ and $\mathbfss{v}_{\alpha}(t)$ denote the position and velocity vector of the $\alpha$-th particle with respect to the proto-halo centre of mass at time $t$. In the following, we will interchangeably refer to ``angular momentum'' as "spin''. This
should not be confused with the \emph{dimensionless spin parameter}, $\lambda$, a scalar quantity that measures the rotational support of compact structures and which we do not study here.

\subsection{Cosmic web identification}
\label{sec:methods:nexus}

We have identified the cosmic web environments using the \nexus{} method \citep{cautunetal2013}, which is an improved version of the Multiscale Morphology Filter introduced in \citet{aragon2007MMF}. To account for the hierarchical character of the cosmic web, \nexus{} employs a scale-space approach to simultaneously identify both prominent as well as tenuous structures. The multiscale approach of \nexus{} makes it ideal for our analysis of filament haloes since the method has two major advantages compared to other filament finders employed in the literature. First, it returns an unbiased and complete sample of filaments, from the thick arteries feeding clusters to the thin filamentary tendrils that criss-cross the underdense regions. Secondly, the method is self-adaptive to the local filament thickness, and, for example, the principal axes of filaments are determined on the scale where the filamentary character of the local matter distribution is most pronounced. For a detailed study of the properties of \nexus{} filaments and of their halo and galaxy population see \citet[][see also \citealt{veenaetal2018,veenaetal2019,hellwingetal2020}]{cautunetal2014}. A detailed comparison of \nexus{} results with other web finders can be found in \citet{libeskind2018}.

The \nexus{} takes as input the total density field that we have obtained by sampling the $z=0$ density field with a $800^3$ grid ($0.5\hMpc$ grid spacing). To calculate the density, we have projected the particle distribution to the grid using the cloud-in-cell interpolation scheme.
In a first step, \nexus{} smooths the input density field using Log-Gaussian filter, a Gaussian smoothing of the density logarithm.
This filter type has been shown to lead to a better identification of the LSS and return more robust environments (see, for instance, Figure 4 and 5 in \citealt{cautunetal2013}). 
To account for the hierarchical nature of the cosmic web, \nexus{} smooths the density on a set of smoothing scales, starting from $0.5$ to $4.0\hMpc$, where each smoothing scale is a factor of $\sqrt{2}$ higher than the previous one. This approach, known as the scale-space formalism, is the key step that allows \nexus{} to identify filaments with a wide range of thicknesses \citep[e.g. see][]{cautunetal2014}.

\begin{figure*}
	\includegraphics[width=2\columnwidth]{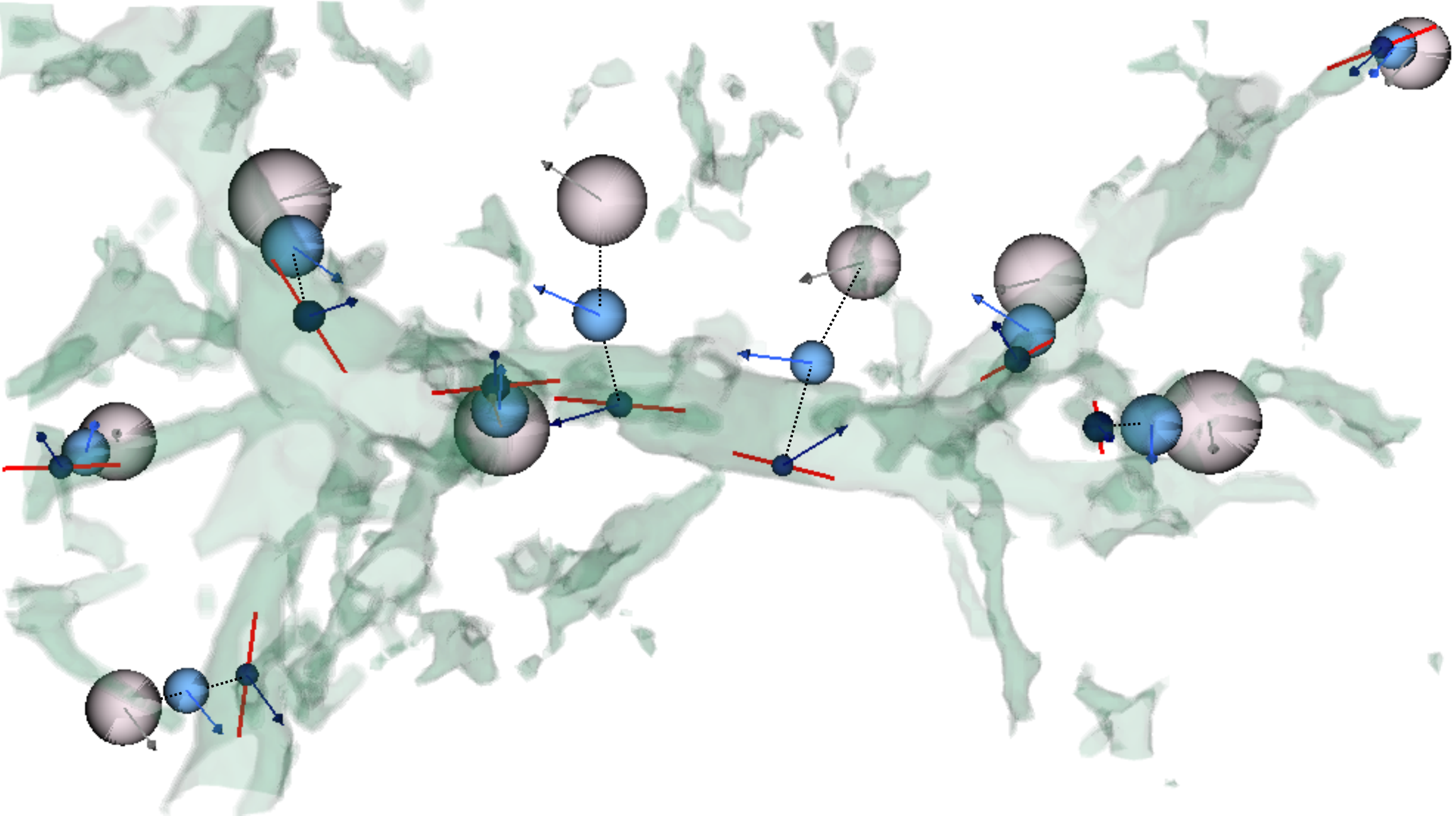}
    \caption{Illustration of the filament distribution (green shaded region) in a small volume of our simulation ($35\times85~ h^{-2}\mathrm{Mpc^2}$ wide and $15~ h^{-1} \mathrm{Mpc}$ thick). The spherical symbols show a subset of filament haloes, with the small dark blue spheres showing their $z=0$ position, and the light blue and white spheres showing the position of the proto-haloes at redshifts $z=1.4$ and $z=80$, respectively. Some of the proto-haloes are connected by a dotted line to better highlight the history of individual objects. The arrows show the spin direction, $\hat{\mathbf{J}}$, at different redshifts, while the red solid line indicates the filament spine at the position of each $z=0$ halo.
    \label{fig:haloes_in_filaments}}
\end{figure*}

For each smoothing scale, \nexus{} calculates the Hessian of the smoothed density field and uses its eigenvalues, $\lambda_1 \leq  \lambda_2 \leq \lambda_3$, to determine an environment signature for each pixel of the input grid. The calculation is rather involved, but qualitatively a pixel has a large filament signature when the matter has collapsed along two directions, i.e. $\lambda_{1} \simeq \lambda_{2} < 0$, and when the variation in density along the third direction is small compared to change along the other two directions, i.e. 
$|\lambda_{1}| \simeq |\lambda_{2}| \gg |\lambda_{3}|$. The three eigenvectors of the Hessian matrix, $\ea$, $\eb$ and $\ec$, determine the principal orientations of the filament and correspond to the first, intermediate and last collapse directions, respectively.

These principal directions are related to the visual appearance of filamentary structures. In effect, filament-like regions are surrounded by a density field with a saddle shaped geometry, and each eigenvector points toward one of the three characteristic directions. The first axis of collapse, $\ea$, indicates the direction of maximum compression of the filament, roughly oriented perpendicular on the plane of the sheet in which the filament is embedded. The orientation of $\ec$ shows the \emph{spine} of the filament: the direction with the least amount of gravitational compression, along which matter flows toward the nodes of the cosmic web. Finally, the axis $\eb$ can be interpreted as the direction that, together with $\ec$, defines the plane of the wall in which the filament is embedded; \eb{} points along the direction of intermediate gravitational influence along which matter flows from the surrounding wall into the filament.

In the last step, \nexus{} combines the environmental signature of all the scales to obtain a scale independent value. This is given by the largest value for a given position and is motivated by the fact that a filament of a given thickness is most easily detectable when smoothing the density with a filter of the same size as the filament thickness. The local filament orientation is given by the eigenvectors corresponding to the smoothing scale with the largest filament signature. All regions whose signature is above a given threshold, which is found by requiring that all resulting filaments are robustly identified, are then classified as filaments.

To each halo we assign the web environment and the web principal axes (i.e. eigenvectors) corresponding to the \nexus{} grid cell in which the halo is located. As we already discussed, we do this step only for the present-day halo distribution and the environment of each proto-halo is that of its $z=0$ descendant. The majority of $z=0$ haloes are found in filaments \citep[e.g. see][]{cautunetal2013,veenaetal2018}, which is why here we study the alignments of filament haloes. 

\section{The evolution of the halo--filament alignment}
\label{sec:alignment_all}

In this section we analyze the evolution of the alignment of the proto-haloes' shape, surrounding tidal field and spin direction with respect to the present-day filaments. The goal is to understand how the proto-halo collapse directions and its angular momentum acquisition are related to the $z=0$ filaments in which the haloes end up. We first start with an overview of our analysis pipeline and a detailed discussion of the motivations behind our choice.

\subsection{Analysis overview} 
\label{sec:alignment_all:overview}

One important difference between our analysis and others in the field is that we correlate proto-haloes with their $z=0$ filaments. As we discussed in the introduction, the main motivation behind this approach is to disentangle the evolution of proto-halo properties from the evolution of their host environments. To better understand this aspect, we illustrate in Figure \ref{fig:haloes_in_filaments} the present-day filaments in a small volume of our simulation. Most of this filamentary structure has been in place since at least $z=2$, with the only difference being that at high redshift the filaments shown in the panel were surrounded by many more tenuous structures \citep[see Figure 21 in][]{cautunetal2014}.

To illustrate the evolution of proto-haloes, Figure \ref{fig:haloes_in_filaments} also shows the positions of a small subset of objects at three redshifts: $z=0,1.4$ and $80$. Here we are interested in depicting the past positions of the proto-haloes relative to the filaments at that time, so the $z=1.4$ and $80$ proto-haloes positions have been corrected for the large-scale displacement that also affects the filaments as a whole. We have achieved this by tracing back in time the positions of all the DM haloes within $10\hMpc$ from each $z=0$ halo, and assuming that this extended mass distribution traces the large-scale peculiar motion. Then, at each redshift, we calculated the proto-halo displacement with respect to the centre of mass of that extend mass distribution.

\begin{figure*}
	\includegraphics[width=\textwidth]{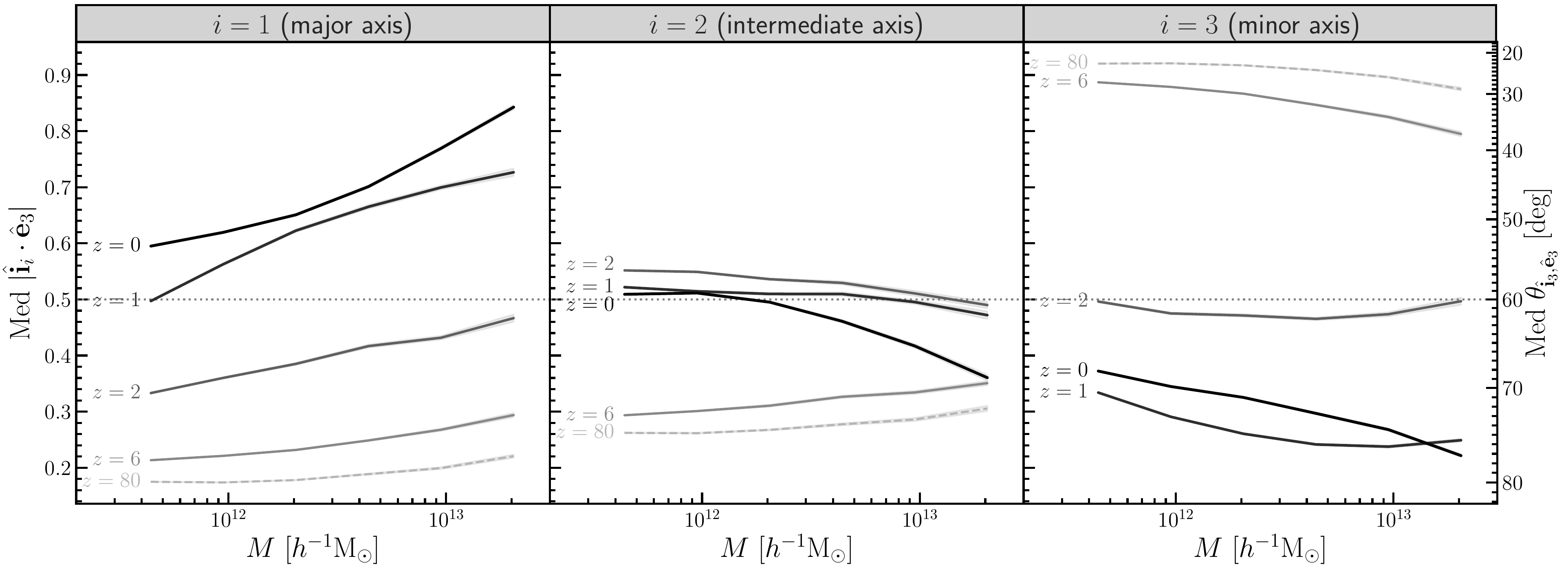}
	\vskip -.2cm
    \caption{Median alignment angle between the major, intermediate, and minor axes of proto-haloes and the spine of their host filament, $\ec$, as a function of halo mass. The curves in each panel correspond to different redshifts, as indicated by the label to the left of them. Each line has associated to it a shaded region that shows the 68 percentile uncertainty in the determination of the median value; this error is typically very small and roughly the width of the line itself, which makes it hard to see. The horizontal dotted line shows the median angle if the halo and filament axes were randomly oriented with respect to each other. For convenience, on the right-side of the figure we give the alignment angle in degrees (this will be given in all other figures too).
    }
    \label{fig:median_alignment_itF}
\end{figure*}

It is worth noting that proto-haloes significantly change their position across time. At high redshift, most objects are typically located in sheet-like environments, where the mass flows mainly from voids, perpendicular to the sheet plane. However, as time passes, filaments and clusters progressively arise in the LSS and proto-haloes start moving mainly along the spine of the filaments \citep{Wang2018a}. Additionally, the principal axes of a given filament can also vary with time. This means that a proto-halo can reside in filaments with different orientations at different times. As pointed in \autoref{introduction}, this makes it hard to interpret the alignment signals within the TTT framework when identifying both haloes and filaments at each redshift. In this work, we prefer to gain insight of the underlying physical processes by reducing the complexity of the problem at hand, which we achieve by keeping fixed as many aspects of the analysis as possible. In particular, we decide to compare the evolution of the haloes' angular momentum with a settled configuration. Hence, we stress that the identification of the cosmic web environments is only performed at $z=0$ and, consequently, the filament axes, $\ea$, $\eb$ and $\ec$, are always fixed to their present-day values.

In summary, Figure \ref{fig:haloes_in_filaments} is useful to visualize the steps we followed to perform the identification of haloes, proto-haloes and filaments:
\begin{enumerate}
    \item Identification of DM haloes by means of a FOF algorithm at present time (dark blue spheres) and computation of their shape and angular momentum. 
    \item Tracking of the haloes' particles back in time and calculation of the properties of the corresponding proto-haloes (light blue and white spheres).
    \item Detection of filaments at present time (light green regions) and their preferential directions (the filament spine at each halo position is shown as a red line in the figure).
\end{enumerate}

To perform our analysis we determine the angle, $\theta$, between the angular momentum of the proto-haloes at different redshifts and the direction of the filaments in which they are embedded at present time. This is implemented by calculating the absolute value of the dot product,
$$
\cos(\theta_{\Jdir,\ei})=|\Jdir \cdot \ei|,
$$ 
between the spin direction, $\Jdir$, and the filament orientation, $\ei$, with $i=1,2,3$. We take the absolute value since filaments have an orientation and not a direction, that is both $\ei$ and $-\ei$ represent filaments with the same orientation. We follow the same procedure to determine the angles between the shape of the (proto-)haloes and the filament orientation, but now calculating the dot product $|\ii \cdot \ei|$, where $\ia$, $\ib$ and $\ic$ represent the eigenvectors of the inertia tensor (see Eq. \ref{eq:IT}).

We are interested in statistically quantifying the spin--filament alignment for different population of haloes in mass and redshift. The ideal way is to study the full probability distribution function (PDF) of the dot product between the spin and the preferential directions of the filaments, and to compare this with the distribution expected for random orientations. However, in many cases the dimensionality of the problem makes it challenging to compare the full PDFs. By studying the shape of such distributions we find, like many previous authors, that the median is in fact sufficient to quantify the degree of alignment. Hence, when analyzing the dependence of the alignment with halo mass and redshift, we decide to characterize the PDFs in terms of their median value. To assess the robustness of these estimates, we calculate 68 percentile uncertainties using bootstrapping. This proceeds by randomly sampling the haloes within a certain mass bin and at a given redshift and calculating the median of each distribution. This process is repeated $50$ times and the mean value and dispersion of the corresponding medians are, respectively, our estimators for the alignment and its uncertainty.

\subsection{Halo shape--filament alignment} 
\label{sec:alignment_all:shape_filament}

Figure \ref{fig:median_alignment_itF} shows the evolution of the alignment between the proto-haloes' shape and the spine of the filament in which they are embedded at present time. The different curves in each panel represent the median angle between one of the axes of the ellipsoid that best fits the halo particles, with $i=1,2,3$ corresponding to the major, intermediate and minor axis, respectively, and the spine of the present-day filament, given by the eigenvector $\ec$. The median angles are shown as a function of the halo masses.

At present-day we find that the dot product between the halo major axis and the filament spine is above 0.5, which indicates that the major axis is preferentially aligned with the filament spine. This alignment is largest for massive haloes, where $|\ia \cdot \ec|\simeq0.8$, and decreases for low mass haloes, in good agreement with previous studies \citep[e.g.][]{aragoncalvoetal2007,Hahn2007,Shao2016,veenaetal2018}. The intermediate axis shows a nearly random alignment with \ec{}, however the halo minor axis is strongly perpendicular on the filament spine, with this trend being stronger for massive haloes. The present-day configuration between the shape of haloes and their host filaments has been explained as a consequence of anisotropic accretion, with most of the newly accreted matter coming into haloes along the filament axis in which they are embedded \citep[e.g.][]{Libeskind2014,Kang2015,Shao2018}.

When studying the alignment at higher redshift, we find a substantial evolution of the orientation of proto-halo shapes. For example, in the initial conditions, the Lagrangian patches that end up collapsing into haloes have major axes that show a strong tendency to be perpendicular to the filament spine. This trend has a rather weak dependence on halo mass. The proto-halo minor axis at $z=80$ is very strongly aligned with the filament spine, with a median value $|\ic \cdot \ec|\simeq0.9$, which indicates that on average there is only a ${\sim}25^\circ$ misalignment between the proto-halo minor axis in the initial conditions and the spine of the filaments.

The strong alignment between the $z=80$ proto-halo shapes and the filaments is largely a manifestation between the alignment in the initial conditions of the proto-halo mass tensor and the local tidal field \citep[e.g.][]{vandeWeygaert1996,porcianietal2002b,Rossi2013}. As we will see in section \ref{sec:tidal_field_vs_filaments}, the preferential axes of the present-day filaments and the initial tidal field are very well aligned. It is the integrated effect of the tidal field from initial conditions to the present time that determines the directions of compression, and thus the shape of the Lagrangian patch that collapses to form a halo.

The various lines in Figure \ref{fig:median_alignment_itF} show how the proto-haloes reorient their shapes as they collapse into the present-day haloes. For example, the major axis orientation changes smoothly from being perpendicular to being aligned with the filament spine, with the flip taking place between $z=1$ and 2 (the exact redshift varies with halo mass). A similar trend is seen for the \ic{}--\ec{} alignment, which changes from parallel to perpendicular orientations. Interestingly, in this case we find that the proto-haloes' minor axes showed a larger degree of perpendicular alignment to \ec{} at $z=1$ than at present time (except for the highest mass haloes).

The late evolution of the \ic{}--\ec{} alignment may reflect a change in the proto-haloes' orientation due to secondary tidal torques exerted by the nearby structure. It could also be related to variations in the shape of the proto-haloes during the final stages of collapse. In this sense, it would be interesting to analyse the evolution of the proto-haloes' internal properties over time (e.g. shape, concentration, velocity dispersion). However, such analysis is beyond the scope of the present paper and it will be addressed in future work \citep[also see e.g.][]{hellwingetal2020}.

\subsection{Alignment between the tidal field and the filaments}
\label{sec:tidal_field_vs_filaments}

\begin{figure}
	\includegraphics[width=\columnwidth]{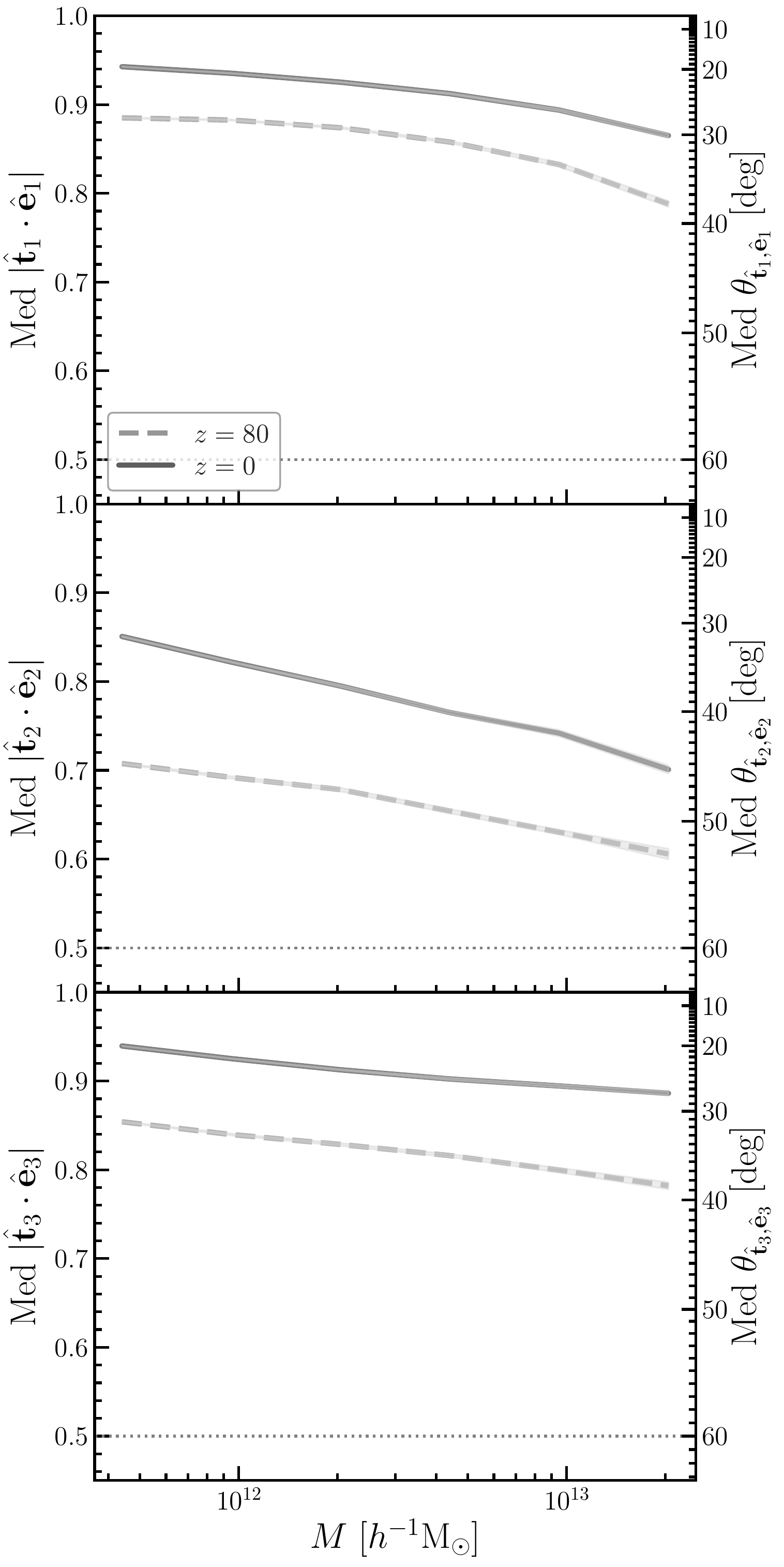}
	\vskip -0.2cm
    \caption{Median alignment between the tidal tensor axes, \ti{}, and the preferential axes of the present-day filaments, \ei{}, with $i=1,2,3$, as a function of halo mass. The tidal tensor is calculated at the centre of mass of each proto-halo at $z=0$ (solid lines) and $z=80$ (dashed lines). The horizontal dotted line shows the expectation if $\ti$ were randomly oriented with respect to the $\ei$.
    }
    \label{fig:median_alignment_tfF}
\end{figure}

One of the key ingredients in the TTT is the orientation of the tidal field. In effect, as explained in section \ref{introduction}, the model states that the angular momentum originates from the misalignment between the shape of a given proto-halo and the tidal field produced by its surrounding density perturbations. These quantities are represented in Equation \eqref{eq:TTT} by the inertia tensor, $I_{lk}$, and the velocity \emph{shear} or tidal tensor, $T_{jl}$. The implementation of the linear approximation in the formulation of the TTT implies that the cross-talk between $I_{lk}$ and $T_{jl}$ does not depend on time, and hence the direction of the angular momentum is fixed in the initial conditions of the proto-haloes' history.

In order to test this assumption, we would like to know how rapidly the inertia tensor and the tidal field change with time (for the evolution in the inertia tensor orientation see section \ref{sec:alignment_all:shape_filament}). Additionally, we are interested in quantifying to what extent the preferential axes of filaments are related to the tidal field. Hence, in this section we analyze the evolution of the halo-centric tidal field with respect to the preferred axes of the filaments.

First, let us remark that the relevant tidal field for spin acquisition is the one defined on scales similar to the size of each proto-halo \citep[e.g.][]{white1984,porcianietal2002a}. To appropriately capture this effect, we employ an adaptive smoothing to calculate the tidal field at each halo position. Here, we study two epochs: the present time and the initial conditions. At the present time, motivated by \citet{Libeskind2014}, we calculate the tidal field using a smoothing scale, $R_\mathrm{h}=4R_{200}$, where $R_{200}$ is the distance from the halo centre that encloses a spherical overdensity of $200$ times the mean background matter density. In the initial conditions, we use a smoothing scale of $R_\mathrm{h}=2R_\mathrm{sph}$, where $R_\mathrm{sph}$ corresponds to the radius of a sphere that encloses the same volume as the halo in Lagrangian space.

\begin{figure*}
	\includegraphics[width=2\columnwidth]{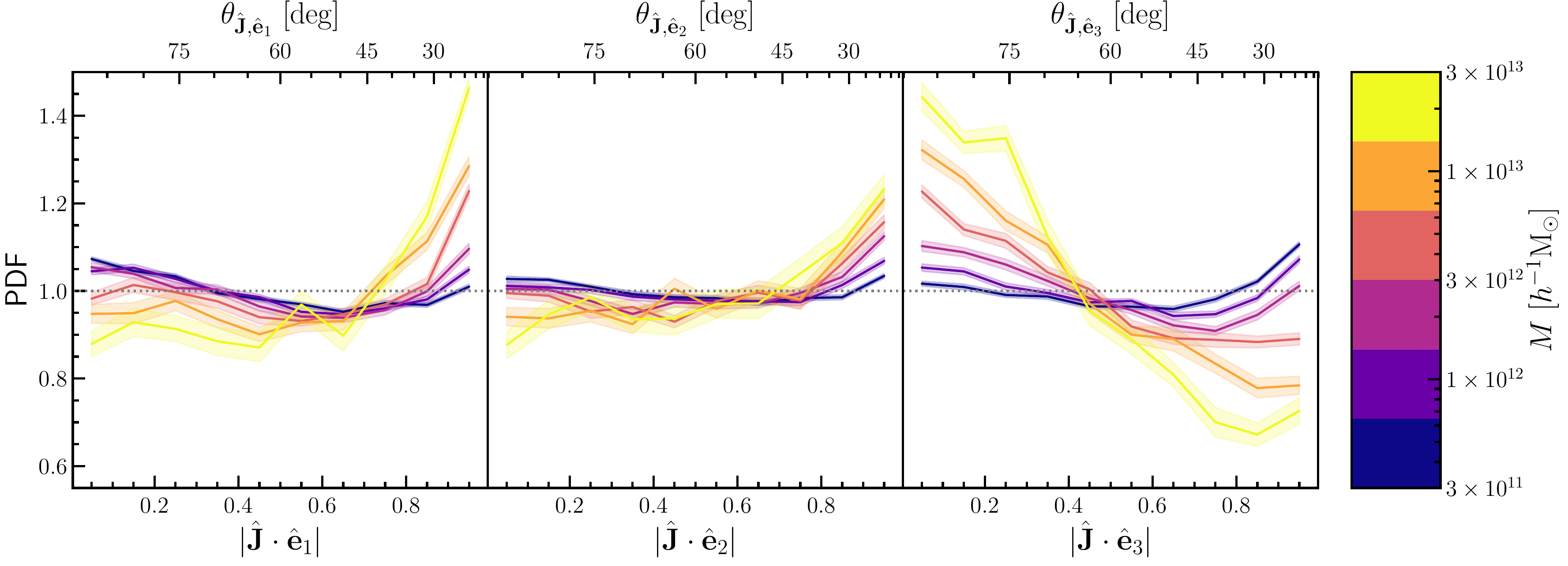}
    \caption{The present-day distribution of the alignment angle between the haloes' spin direction, $\Jdir$, and the preferential axes, $\ea,\eb$, and $\ec$, of the filaments in which the haloes are embedded. The various colors correspond to haloes of different masses, as shown by the right-hand side color bar. The horizontal dotted lines show the null-alignment expectation, i.e. if the haloes' spin would be randomly oriented with respect to their host filaments, while the shaded area around each curve shows the 68 confidence interval with which the PDF can be distinguished from the random distribution given the limited number of haloes.}
    \label{fig:mass_alignment_dist_1x3}
\end{figure*}

To obtain the tidal field, we first calculate the density field, $\delta(\mathbf{x})$, using cloud-in-cell interpolation on a regular grid. In the next step, we invert the Poisson equation, $\nabla^2\phi=\delta$, in Fourier space in order to calculate the gravitational potential $\phi$. Then, we estimate the tidal tensor as
$$
T_{jl}=\frac{\partial^2 \phi}{\partial x^j\partial x^l} \; .
$$
After this, we smooth the tidal field using several Gaussian filters, each one with a characteristic scale $R_\mathrm{G}$. Hence, we generate a set of measurements $T_{jl}(\mathbf{x},R_{\mathrm{G}})$ of the tidal field over a 3D spatial grid combined with a grid of smoothing scales, $R_\mathrm{G}$. Then, we assign one of these measurements to each halo by finding the cell of the tidal field grid that best matches the halo's centre of mass and its characteristic smoothing scale, $R_h$. Finally, we diagonalize this halo-centric tidal tensor and thus obtain the corresponding eigenvalues $\lambda_i$ and eigenvectors \ti{}, with $i=1,2,3$. 

In Figure \ref{fig:median_alignment_tfF} we show the median alignment between the eigenvectors of the tidal field, \ti{} and the preferred filament axes as a function of halo mass. In order to analyze how this alignment varies with time, we present measurements of the tidal field both at present time and at the initial conditions.

We can see that there is a clear correlation between the filament axes and the preferred directions of the tidal field. In effect, at the present time (solid lines), more than $50\%$ of the haloes are located within filaments whose first axis of collapse, \ea{}, is oriented less than $30$ deg from the major axis of the surrounding tidal field, and this alignment is even stronger between the spine of the filaments and \tc{}. Although slightly less correlated, the alignment between \eb{} and \tb{} is also significant. This is to be expected \citep{vanhaarlem&vandeweygaert1993,vandeWeygaert1996,weygaert&bond2008}, since the cosmic filaments are actually shaped by the large-scale tidal fields. Furthermore, the filaments and the halo-centric tidal field are identified on different scales, so it should not be surprising that they are not perfectly aligned.

When we look back in time and observe the preferred directions of the halo-centric tidal tensor at the initial conditions (dashed lines), we find that there already exists a strong correlation with the orientation of the present-day filaments, even at such high redshift. The median alignment signals show that the vast majority of proto-haloes are located in regions whose surrounding tidal field is oriented in a similar way to the filaments to be formed there. Hence, these results indicate that the orientation of the axes \ti{} remains largely unchanged over time. If anything, there is a slight evolution towards configurations more aligned with the filaments. Again, this is to be expected, since the solid curves correspond to highly correlated structures identified at the same redshift.

The tight alignment between the tidal field orientation and the filament axes, especially with respect to \ea{} and \ec{}, indicates that it is justified to use the preferential axes of the present-day filaments as a proxy for the preferred directions of the large-scale tidal field. On the other hand, the fact that we see only small changes with time in the orientation of the tidal field suggests that deviations from TTT are more likely to be caused by the large evolution in the orientation of proto-halo shapes (see section \ref{sec:alignment_all:shape_filament}), i.e. in the non-linear evolution of the inertia tensor $I_{lk}$. 

\subsection{Halo spin orientation at present-day}
\label{sec:alignment_all:spin_filament_present-day}

We now continue to the main topic of our analysis, the alignment between the angular momentum of haloes and the axes of their host filaments. To this end, we will start with the present-day configuration. In Figure \ref{fig:mass_alignment_dist_1x3} we show the PDF of $\Jdir\cdot\ei$ at the present time, with $i=1,2,3$. The different colored lines show the dependence of the alignment signal with halo mass, as shown in the color bar on the right-hand side.

For all the panels in Figure \ref{fig:mass_alignment_dist_1x3} we find that the $\Jdir\cdot\ei$ PDF shows only small deviations from a uniform distribution, which is shown via the horizontal dotted line. This indicates that in general the halo spin has only a weak tendency to be aligned with the filament axes and that the signal we study is a small excess of parallel (i.e. the majority of haloes have $\Jdir\cdot\ei > 0.5$) or perpendicular (i.e. the majority of haloes have $\Jdir\cdot\ei < 0.5$) orientations.

The right-most panel in Figure \ref{fig:mass_alignment_dist_1x3} shows the alignment with respect to $\ec$, the spine of the filament. We find that the halo spin--\ec{} alignment varies with halo mass, from a small excess of parallel configurations for low-mass haloes to a preponderance for perpendicular orientations for high-mass haloes. This aspect of the alignment between halo spin and the LSS has been widely explored in the past years \citep[e.g.][]{aragoncalvoetal2007,pazetal2008,codisetal2012,libeskindetal2013,foreroromeroetal2014,lopezetal2019}, and has been explained as a manifestation of anisotropic secondary accretion \citep[e.g.][]{veenaetal2018,Wang2018a}. The high mass population is dominated by haloes that are still vigorously accreting material along the spine of their host filament, which leads to a net increase of angular momentum along a direction perpendicular to the infall one.

A theoretical framework for the \Jdir-\ec{} alignment has been proposed by \citet{laigleetal2015} and \citet{codisetal2015}, who have showed that this trend follows from the vorticity pattern produced by an implementation of TTT constrained to filamentary regions. This vorticity field is octupolar and oriented along the filament spine in the vicinity of a saddle point, and becomes perpendicular away from it. Low-mass haloes typically form in one of the octants, and thus acquire angular momentum along the filament. High-mass haloes, on the other hand, frequently overlap adjacent octants with opposite spin directions and hence end up acquiring angular momentum preferentially perpendicular to their host filament. Moreover, high-mass haloes are usually located away from the saddle points, which enhances the effect. 

However, in Figure \ref{fig:mass_alignment_dist_1x3}, although the alignment signal systematically inverts towards lower masses, the probability of finding haloes with angular momentum perpendicular to the spine of their host filament does never fall below the random expectation. More concretely, haloes with $M<10^{12}\hMsun$ slightly overpass the horizontal line that defines a random distribution at the right-hand extreme (i.e. parallel configurations), as expected, but also at the left-hand extreme (i.e. perpendicular configurations). This subtle bimodality suggests that these distributions could be mixing populations of low-mass haloes that have suffered different processes during their evolution. In the next section we will show that, indeed, this bimodality can be partially decoupled when we take into account how DM haloes deviate from the TTT predictions.

These trends can be further analyzed by looking at the orientation of the angular momentum with respect to $\ea$ and $\eb$, which is shown in the left-hand and the central panels of Figure \ref{fig:mass_alignment_dist_1x3}, respectively. These axes determine the plane perpendicular to the filament spine, \ec{}. For high-mass haloes, \Jdir{} is preferentially perpendicular on \ec{} and thus the spin lies within the $(\ea,\eb)$ plane, where it shows a preference to be oriented along \ea{}, the first direction of collapse. In other words, the spin of high-mass haloes is mainly along the normal to the large-scale wall that surrounds a halo's host filament. Low-mass haloes show a different alignment, with their spin orientations showing a mild preference of being perpendicular to \ea{}.

\begin{figure}
	\includegraphics[width=\columnwidth]{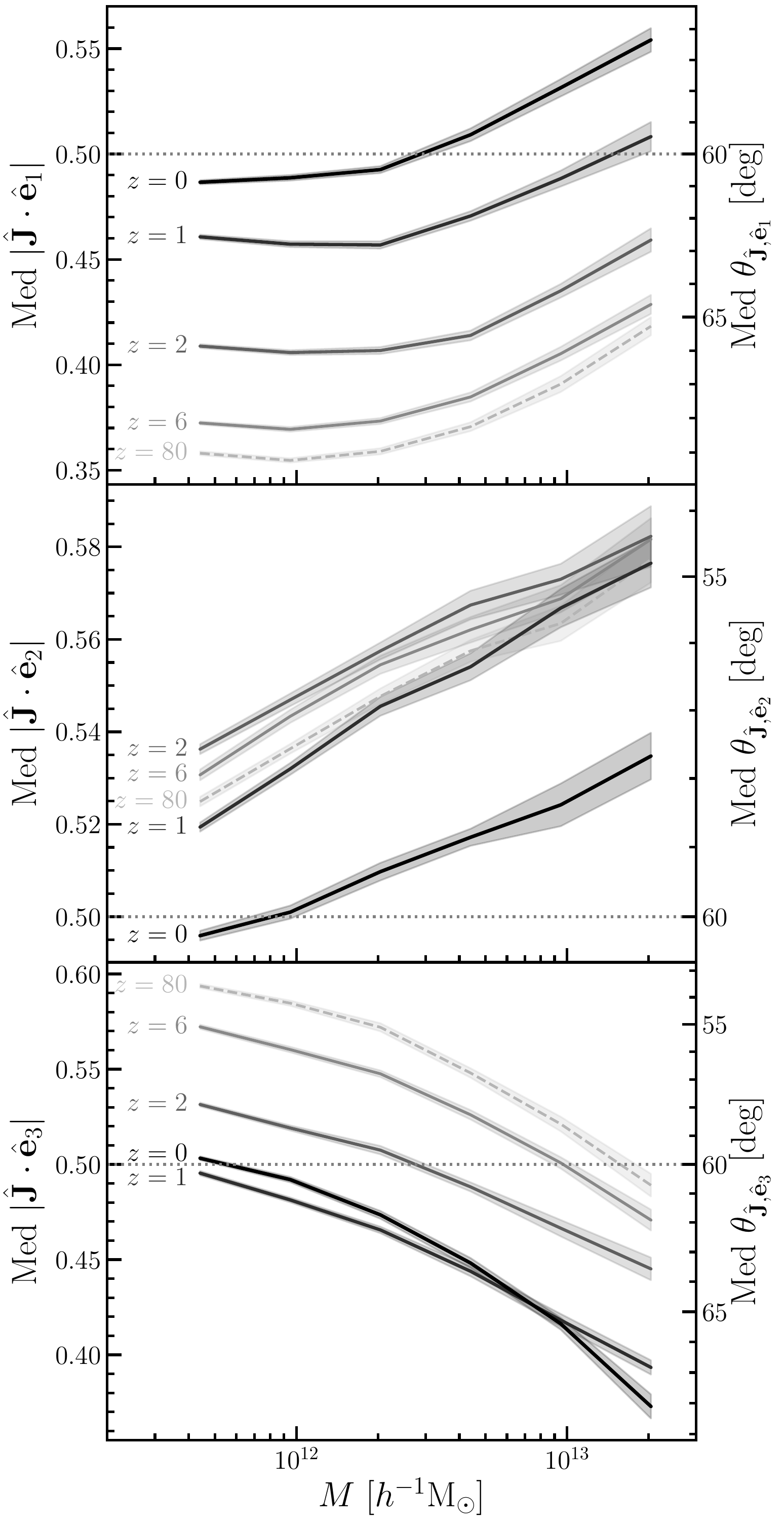}
	\vskip -.3cm
	\caption{The evolution of the median alignment between a halo's angular momentum and the axes of its host filament as a function of halo mass, $M$. Each panel corresponds to the alignment with a different filament axis, that is from top to bottom: $\ea$, $\eb$ and $\ec$. Each curve represents the median alignment between the halo spin at that redshift and the present-day, $z=0$, filament axes. Shaded areas show the $68\%$ uncertainty in the determination of the median alignment angle for each mass bin.
	}
    \label{fig:median_alignment_3x1}
\end{figure}

The variation of the halo spin--filament alignment with halo mass is best summarized in Figure \ref{fig:median_alignment_3x1}, where we show the median alignment angle for haloes in narrow mass bins. Focusing on the results in the bottom panel of the Figure (\Jdir{}--\ec{} alignment), and looking particularly at the curve corresponding to $z=0$, we find that the majority of halo masses studied here have median values $|\Jdir\cdot\ec|<0.5$, indicating a preference for perpendicular orientations with respect to the spine of the filament. Only haloes less massive than $5\times10^{11}\hMsun$ have an excess of parallel orientations (i.e. median $|\Jdir\cdot\ec|>0.5$). This is in good agreement with previous works \citep[e.g.][]{aragoncalvoetal2007,hahnetal2010,codisetal2012,foreroromeroetal2014}, although the exact transition mass from an excess of parallel orientations to an excess of perpendicular ones depends sensitively on the web identification method and can vary by an order of magnitude between various web finders. This variation is driven by the dependence of the spin--filament alignment on the properties of filaments \citep[e.g.][]{aragon2014,veenaetal2018,veenaetal2020}, with each web finder identifying somewhat different filament populations \citep{libeskind2018}. For the median spin alignment with the other two filament axes, \ea{} and \eb{}, see the top and middle panels in Figure \ref{fig:median_alignment_3x1}.

\subsection{Time evolution of halo spin orientation}
\label{sec:alignment_all:spin_filament_evolution}

In order to better understand how the present-day configuration arises, in Figure \ref{fig:median_alignment_3x1} we show how the alignment between the angular momentum and the filament axes evolves with time. To this end, the different curves in each panel represent the median of $|\Jdir\cdot\ei|$ at different redshifts, while the horizontal dotted lines at $|\Jdir\cdot\ei|=0.5$ correspond to the median expectation for randomly oriented spins, i.e. the threshold between preferentially parallel and preferentially perpendicular alignment. 

We first notice that the spin direction of proto-haloes in the initial conditions is not randomly oriented with respect to the present-day filaments. Rather, it seems to be preferentially perpendicular to the first axis of collapse, $\ea$, and mostly along $\eb$ and $\ec$, with some dependence on halo mass. For example, the median $|\Jdir\cdot\ea|$ value increases with halo mass, indicating that a higher fraction of low-mass haloes have spins perpendicular to \ea{} than high-mass haloes. The median $|\Jdir\cdot\eb|$ alignment also increase with halo mass, but, since its values are above 0.5, i.e. the expectation for random orientations, the interpretation is that a lower fraction of low-mass haloes have spins along \eb{} than their high-mass counterparts. Finally, $|\Jdir\cdot\ec|$ decreases with halo mass, with the highest-mass haloes shown in the figure having spins  whose directions are close to random with respect to the filament spine.

As we show in section \ref{sec:tidal_field_vs_filaments}, the preferential axes of the initial tidal field, \ta{}, \tb{} and \tc{}, are well aligned with the axes of the present-day filaments, \ea{}, \eb{}, and \ec{}, respectively. This means that the alignment between the initial spin and the filament axes is largely the same as the alignment with the initial tidal field. Hence, these results are in good agreement with \citet{porcianietal2002b}, who noticed that, given the strong correlation in the initial conditions between the proto-halo shape and the tidal tensor, TTT predicts that the angular momentum is preferentially perpendicular to \ta{} and mostly aligned with \tb{} and \tc{}.

Figure \ref{fig:median_alignment_3x1} illustrates that the initial conditions and the present-day proto-halo spins are oriented differently with respect to filaments. This means that TTT does not fully explain the $z=0$ spin--filament alignment and that  late-time non-linear spin acquisition processes play an important role. These non-linear processes systematically increase the median $|\Jdir\cdot\ea|$ values, that is the spin reorients to point more closely along the axis of first collapse, \ea{} (uppermost panel). Since initially \Jdir{} was preferentially perpendicular to \ea{}, the spin reorientation leads to a present-day spin that is only marginally perpendicular to \ea{} for low-mass haloes and marginally aligned with \ea{} for high-mass haloes. This observation raises an interesting question: does the non-linear evolution proceeds to erase the \Jdir--\ea{} initial alignment? Or is this result just a coincidence, and if we were to follow the haloes into the future we would find that $|\Jdir\cdot\ea|$ keeps growing even higher (i.e. \Jdir{} reorients to become even more parallel to \ea)?

The evolution of the spin direction with respect to \eb{} (central panel in Figure \ref{fig:median_alignment_3x1}) is considerably different. At redshift $z\geq2$, non-linear processes hardly affect the \Jdir--\eb{} alignment, which shows a very mild increase with time. However, the trend reverses at lower redshift, where the excess of parallel \Jdir--\eb{} configurations decreases with time. In particular, the change from $z=1$ to present-day is characterized by a sudden jump towards nearly random \Jdir--\eb{} configurations.

\begin{figure}
	\includegraphics[width=\columnwidth]{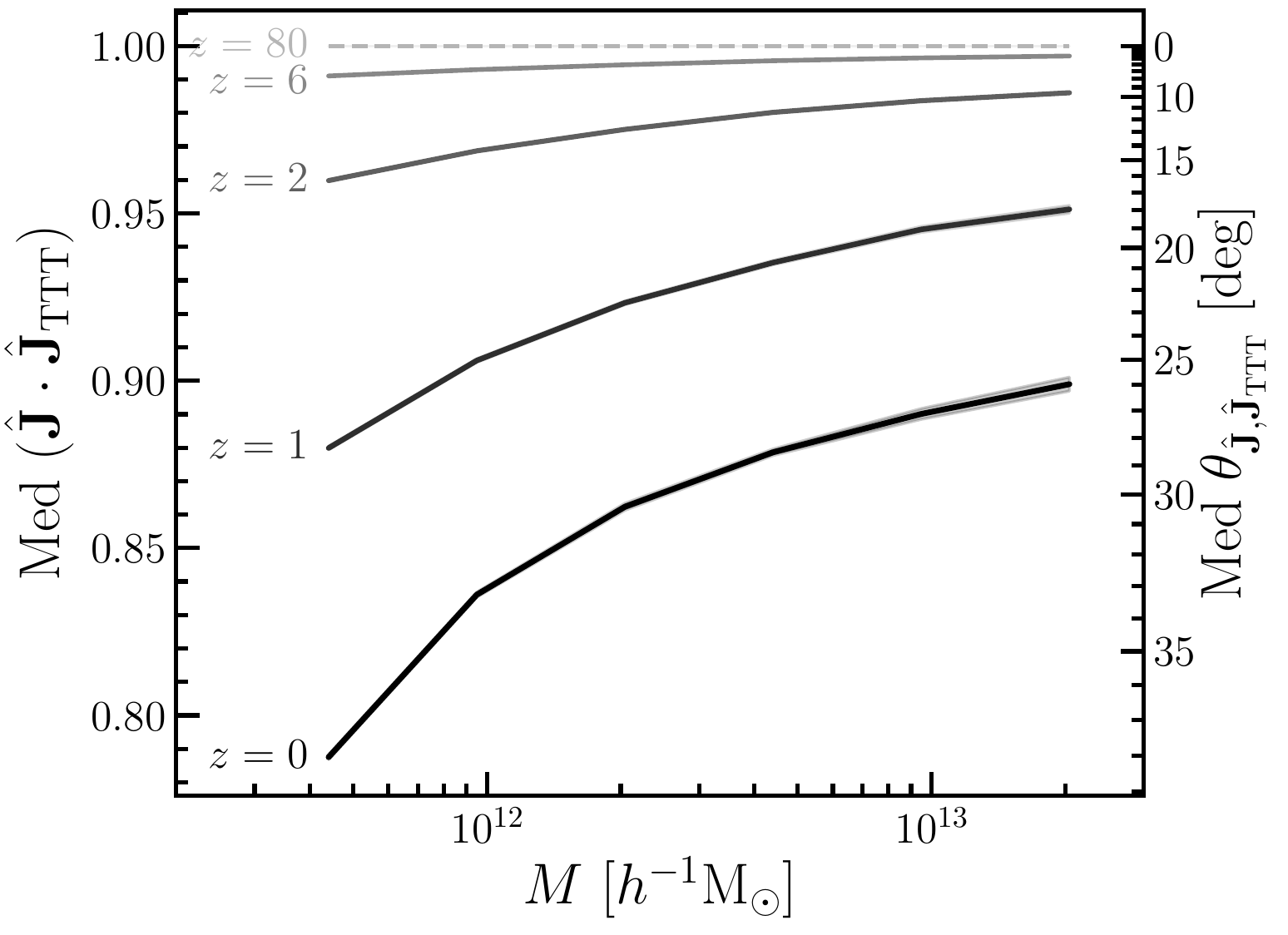}
	\vskip -.3cm
	\caption{Median alignment between the haloes spin at different redshifts and the direction predicted by the TTT, i.e. the proto-halo spin in the initial conditions. The median is shown as a function of halo mass, and the shaded areas show the $68\%$ confidence interval for each mass bin (this is thinner than the curves themselves).
	}
    \label{fig:median_autoalignment}
\end{figure}

With respect to $\ec$ (lower panel in Figure \ref{fig:median_alignment_3x1}), the angular momentum shows a clear evolution that goes from mostly aligned at high redshift to a preference for perpendicular configurations towards the present time. This evolution progresses rapidly until $z\sim1$, after which the median of  $|\Jdir\cdot\ec|$ remains roughly constant. However, it is interesting that, at this point, high mass haloes seem to continue with the high redshift tendency of becoming more perpendicular to the spine of the filament, whereas low mass haloes actually reverse this trend. 

In general, we find that the spin alignment with each filament axis has changed by roughly the same amount for all halo masses. This suggests that the $|\Jdir\cdot\ei|$ trend with halo mass seen at $z=0$ is not due to non-linear spin acquisition processes, but is actually set in the initial conditions, and it reflects that the orientation of the initial proto-halo spin depends on its mass, as predicted by the anisotropic TTT formalism of \citet{codisetal2015}. Overall, the non-linear evolution has lead to modest changes in alignment angle. The largest variation is in the \Jdir{}--\ea{} alignment, which has changed on average by 9 deg. The alignment with the other two filament axes, \eb{} and \ec{}, has changed by 3 and 7 deg, respectively. These results suggest that, to a first approximation, the change in spin orientation can be described, especially for $z\geq1$, as a rotation around \eb{}, which is the filament axis that sees the least amount of change in spin alignment.

We now proceed to study how the change in spin--filament alignment compares to the overall change in proto-halo spin orientation. For this, we calculate the alignment between the spin in the initial conditions and the spin at later redshifts. We remind the reader that we limit this calculation to proto-haloes that at present-day end up in filaments, since they are the topic of our research. The median value of this alignment at different redshifts is shown in Figure \ref{fig:median_autoalignment}, where we plot it as a function of halo mass.

We find that, as we approach present-day, the proto-halo spin shows a larger misalignment compared to its TTT prediction. At $z=0$, the median misalignment angle ranges from $\sim 35$ deg at low masses to $\sim 25$ deg at high masses. Our results are consistent with \citet{neyrincketal2019} and are slightly below those of \citet{porcianietal2002a}, who found a $\sim 40$ deg median misalignment angle \citep[to see a disc][]{}.

Figure \ref{fig:median_autoalignment} shows that non-linear effects reorient on average the proto-halo spin by ${\sim}30$ deg. This is more than a factor of three times higher than the change in spin--filament alignment angle seen in Figure \ref{fig:median_alignment_3x1} and indicates that non-linear spin growth leads to spin direction changes that are largely (but not completely) uncorrelated to the filament axes.

\section{The correlation between spin growth and spin--filament alignment}
\label{sec:alignment_growth}

\begin{figure*}
	\includegraphics[width=2\columnwidth]{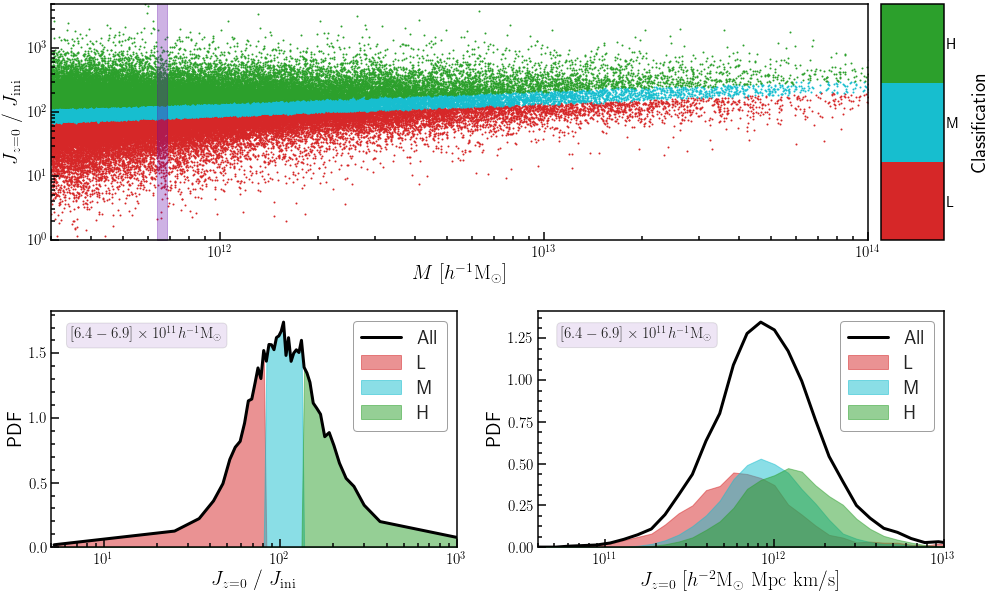}
    \caption{ \textit{Top panel:} the halo net angular momentum growth (i.e. the present-day angular momentum divided by its value in the initial conditions) as a function of halo mass. The green, blue, and red points indicate the \Wh{}, \Mh{}, and \Lh{} subsamples, and correspond to the halo terciles (defined within narrow mass bins) with the highest, medium, and lowest net angular momentum growth, respectively.
    The red and green points represent the \Lh{} and \Wh{} samples, i.e. haloes that have acquired angular momentum below or above the TTT expectations, respectively. The light blue points show the \Mh{} sample, i.e. haloes whose angular momentum has grown as expected from the model. \textit{Bottom left:} the distribution of the net angular momentum growth for the halo mass bin highlighted in purple in the top panel. We show the PDF for all haloes (black line) and for haloes in each of the three subsamples (coloured areas). The height of the coloured areas has been re-scaled to better appreciate the relation with the parent sample: inside narrow mass bins, our classification corresponds to the terciles of the distribution of net angular momentum growth. \textit{Bottom right:} same as the left-hand plot, but for the present-day angular momentum.
    }
    \label{fig:gp_classification}
\end{figure*}

Our main goal is to understand the spin--filament alignment within the TTT framework. An intriguing question is to what extent this alignment is correlated to deviations in spin growth from the TTT prediction? For example, it is conceivable that haloes whose spin growth followed TTT for longer, i.e. until low redshift, could have a spin--filament alignment that more closely resembles the alignment present in the initial conditions.
Here, we investigate if the spin--filament alignment is indeed correlated to halo spin growth. As we shall see shortly, this is an essential ingredient to understand the evolution of the halo spin--filament alignment.

Another motivation is provided by recent studies that have pointed out an interesting correlation between the magnitude of the halo angular momentum and its orientation with respect to filaments. For example, \citet{veenaetal2020} have shown that massive haloes whose spin is perpendicular to the filament spine rotate on average faster (i.e. have higher spins) than haloes with spins along the filament spine. A similar effect has been pointed out by \citet{lopezetal2019}, who have shown that the haloes which experienced the largest amount of spin growth are also more likely to have spin orientations perpendicular to the filaments surrounding them.

\subsection{Net angular momentum growth of haloes}
\label{sec:alignment_growth:net_spin_growth}

Here, we follow the \citet{lopezetal2019} approach and define the \textit{net spin growth} as the ratio between the present-day spin, $\mathbf{J}_{z=0}$, and the proto-halo spin, $\mathbf{J}_{\rm ini}$, in the initial conditions, which we take as the matter distribution at redshift, $z=80$. This definition is motivated by the TTT framework, in which the angular momentum at some time is the initial spin times a time-dependent growth factor (see Eq. \ref{eq:TTT}). Thus, within the TTT approach, the net spin growth is basically the above growth factor.  \citeauthor{lopezetal2019} have shown that haloes with high $\mathbf{J}_{z=0}/\mathbf{J}_{\rm ini}$ ratios for their mass are more likely to have collapsed later and thus followed the TTT predictions for longer. This observation holds for the population as a whole, however not necessarily for individual haloes which could have a high $\mathbf{J}_{z=0}/\mathbf{J}_{\rm ini}$ ratio due to non-linear effects boosting their late-time spin growth.

\begin{figure}
	\includegraphics[width=\columnwidth]{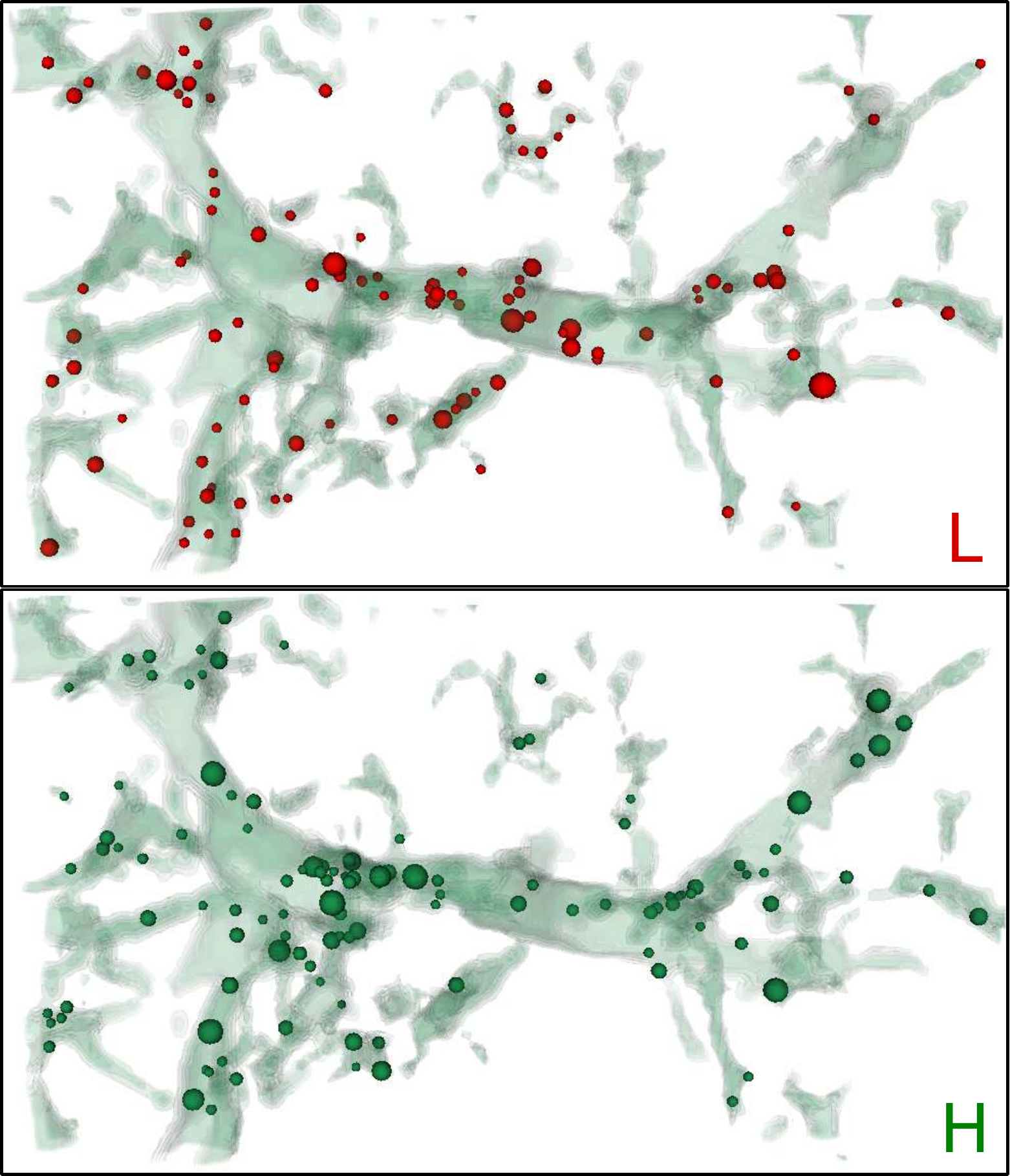}
    \caption{Illustration of the present-day distribution of DM haloes according to their net spin growth. The top and bottom panels show, respectively, the position of \Lh{} and \Wh{} haloes, i.e. haloes whose angular momentum has grown below or above the median TTT expectation (see Figure \ref{fig:gp_classification}). Each halo is represented with a spherical symbol whose size is proportional to the halo mass. The volume and filamentary structure represented here is the same as in Figure \ref{fig:haloes_in_filaments}. For clarity, we show only haloes with masses between $0.3$ and $7\times10^{12}\hMsun$.
    }
    \label{fig:haloes_in_filaments_LMW}
\end{figure}

To keep the analysis simple, we follow the \citeauthor{lopezetal2019} classification and for each narrow bin in mass we divide the halo population into three subsamples corresponding to the haloes with: \textit{highest}, \textit{medium}, and \textit{lowest} net spin growth. We abbreviate these subsamples as \Wh{}, \Mh{}, and \Lh{}, respectively. Each category corresponds to exactly a third of the haloes in each halo mass bin. This nomenclature is slightly different from the one used in \citet{lopezetal2019}, where the same categories are referred as \textit{winner} (W), \textit{median} (M) and \textit{loser} (L). Here, we decide to change the names of the samples (and the abbreviation of the one with the highest net spin growth, W$\rightarrow$\Wh{}) in order to better highlight their properties and the motivation of the classification.

\begin{figure*}
	\includegraphics[width=2\columnwidth]{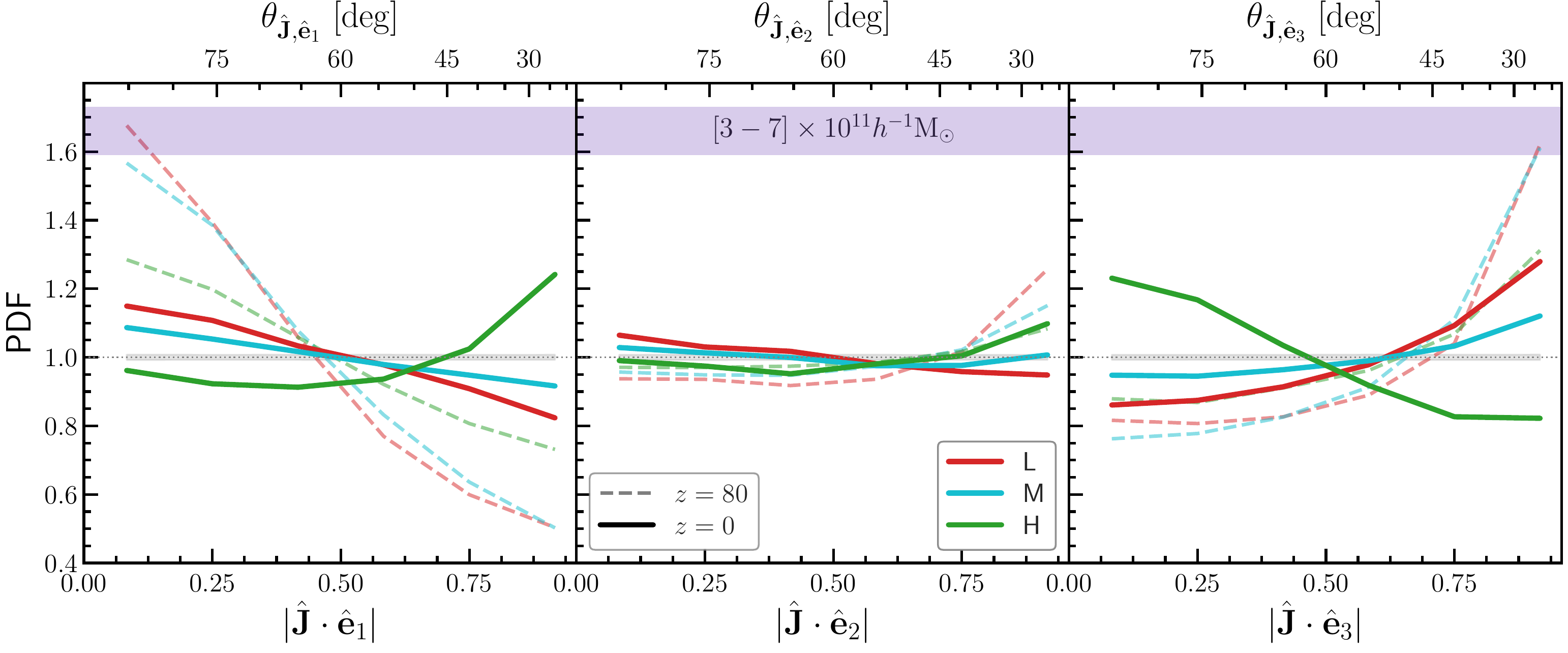}
	\vskip -.15cm
    \caption{Distribution of the alignment between proto-halo spin, \Jdir, and the preferential axes of the present-day filaments, \ei, for haloes with masses between $3$ and $7\times10^{11}\hMsun$. We show both the PDF for the spin at the present time (solid lines) and at the initial conditions (dashed lines). The red, light blue and green curves show the distribution of $|\Jdir\cdot\ei|$ for \Lh{}, \Mh{} and \Wh{} haloes, respectively. The horizontal dotted line represents a random distribution, while the shaded area shows the typical 68 percentile uncertainty with which it can be measured given the limited number of haloes.
    }
    \label{fig:gp_alignment_dist_1x3}
\end{figure*}

This analysis is illustrated in Figure \ref{fig:gp_classification}, where the top panel show the net spin growth, $\mathbf{J}_{z=0}/\mathbf{J}_{\rm ini}$, for individual haloes as a function of their mass. It shows that the $\mathbf{J}_{z=0}/\mathbf{J}_{\rm ini}$ ratio increases systematically with halo mass and illustrates why the \Wh{}, \Mh{}, and \Lh{} subsamples need to be selected using narrow mass bins. The selection is further depicted in the bottom-left panel in Figure \ref{fig:gp_classification}, which shows the distribution of net spin growth for a narrow mass bin. While most haloes have a narrow range in $\mathbf{J}_{z=0}/\mathbf{J}_{\rm ini}$ values, which for the given mass bin cluster around $\mathbf{J}_{z=0}/\mathbf{J}_{\rm ini}{\sim}10^2$, the distribution has extended wings with a considerable fraction of haloes with ratios an order of magnitude higher or lower. The three subsamples, each containing a third of the haloes, are shown in the panel by the three shaded regions.

It is worthwhile to mention that, while the \Wh{} sample corresponds to haloes with the largest net increase in halo spin, it does not necessarily mean that those haloes also have the highest angular momentum for their mass. This is studied in the bottom-right panel of Figure \ref{fig:gp_classification}, where we show the $\mathbf{J}_{z=0}$ distribution for the same narrow mass bin as in the bottom-left panel. While the \Wh{} haloes have on average somewhat higher $z=0$ spins, the distribution largely overlaps with that of the \Lh{} and \Mh{} subsamples. This is because $\mathbf{J}_{\rm ini}$ varies from halo-to-halo, and thus a halo with a high value of net spin growth can end up with a low angular momentum if $\mathbf{J}_{\rm ini}$ was low to start with.

\citet{lopezetal2019} have shown that haloes cluster differently depending on their net spin growth. For halo masses, $M\gtrsim 5\times10^{12}\hMsun$, the \Wh{} sample is typically more clustered, while at lower masses the opposite is true, with the \Lh{} sample being more clustered. To gain insights on the connection between net spin growth of haloes and the filamentary network, we illustrate in Figure \ref{fig:haloes_in_filaments_LMW} the distribution of \Wh{} and \Lh{} haloes superimposed on the filamentary structure in a typical region of our simulation volume. The figure highlights a few subtle differences between the spatial distribution of \Wh{} and \Lh{} haloes. High density regions are typically inhabited by high mass haloes from the \Wh{} sample, which are consequently more clustered and concentrated near the nodes of the cosmic web. Conversely, \Lh{} haloes of lower mass are usually distributed along the filaments and other regions of intermediate density.

\subsection{The $z=0$ spin--filament alignment for \Wh, \Mh, and \Lh{} haloes}
\label{sec:alignment_growth:spin_filament_present-day}

\begin{figure}
	\includegraphics[width=\columnwidth]{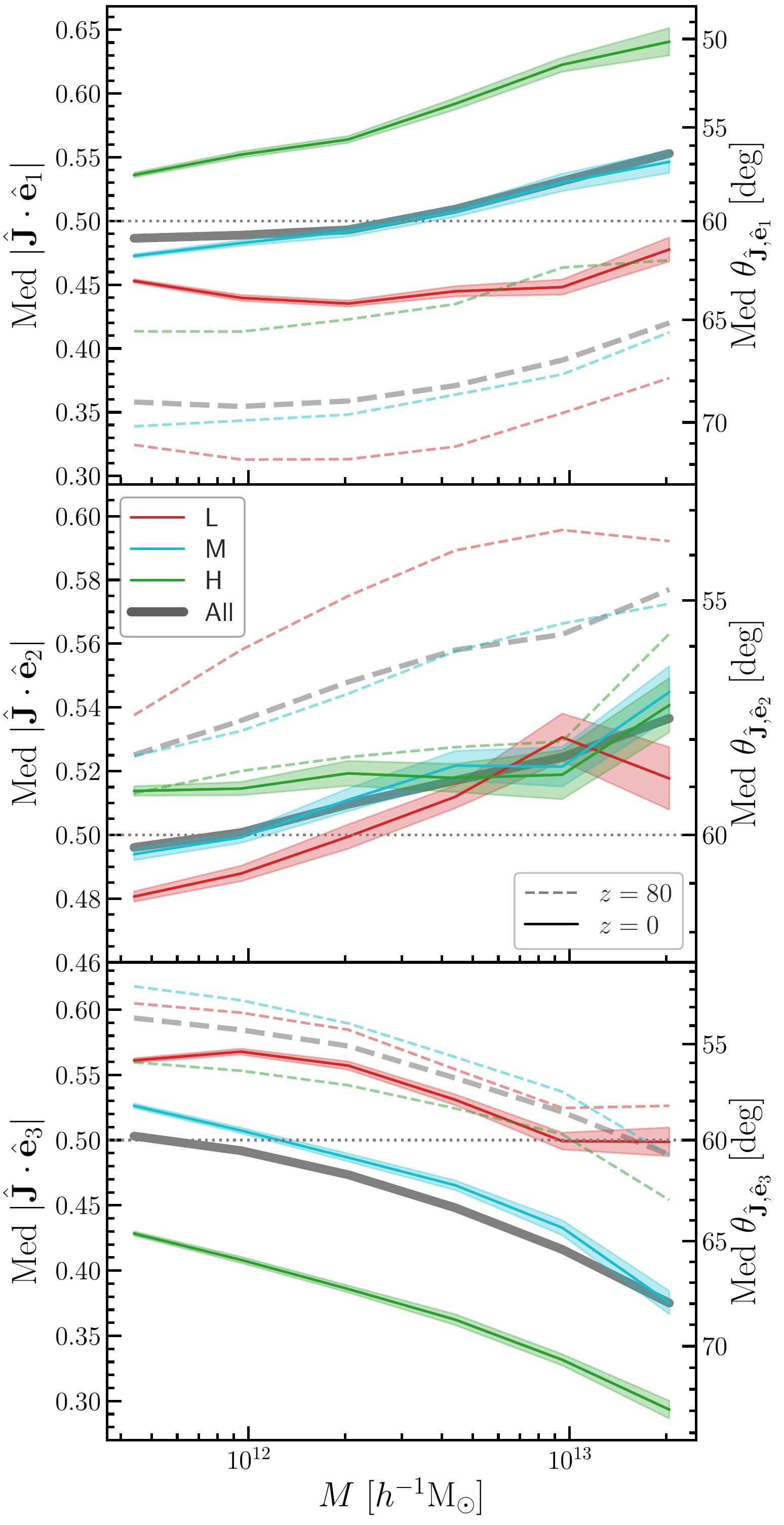}
	\vskip -.3cm
    \caption{Median alignment between the halo spin and the axes of their host filament as a function of halo mass. Each panel corresponds to a different axis, where dashed and solid lines show, respectively, the medians computed using the angular momentum directions at the initial conditions and at present time. The filament axes directions are always fixed at present time. The black curves show the alignment for all haloes (which corresponds to the $z=80$ and $z=0$ curves in Figure \ref{fig:median_alignment_3x1}), whereas the red, light blue and green curves correspond to \Lh{}, \Mh{} and \Wh{} haloes, respectively. Shaded areas show the bootstrap $68\%$ confidence interval for each mass bin.
    }
    \label{fig:gp_median_alignment_3x1}
\end{figure}

We now reexamine the alignment of the haloes' spin with respect to their host filaments in light of our classification according to their net spin growth. We start by illustrating in Figure \ref{fig:gp_alignment_dist_1x3} the PDF of $|\Jdir\cdot\ei|$, with $i=1,2,3$, for haloes with masses between $3$ and $7\times10^{11}\hMsun$, i.e. the low mass end of our halo sample.

Figure \ref{fig:gp_alignment_dist_1x3} reveals a clear correlation between the orientation of $\Jdir$ and the haloes' spin growth. In effect, \Lh{} and \Wh{} haloes show essentially opposite trends of alignment with respect to the preferential axes of the filaments. The spin of \Wh{} haloes is typically oriented along the first axis of collapse, \ea{}, while \Lh{} haloes rotate preferentially along the spine of the filaments, \ec{} (i.e. last collapse axis). This can be most clearly seen in the right-hand panel: the PDF of both the \Lh{} and the \Mh{} samples peak at their right-most extreme (i.e. parallel configurations), while the PDF of \Wh{} haloes has a clear peak in the opposite extreme (i.e. perpendicular configurations). 

The results shown in Figure \ref{fig:gp_alignment_dist_1x3} help us understand a puzzling feature seen in the spin--filament alignment PDF of the full sample of haloes in that mass range, which shows a subtle bimodality (see purple curve in the right-most panel of Figure \ref{fig:mass_alignment_dist_1x3}): it has an excess of both parallel and perpendicular alignments. This is the outcome of combining the  samples \Lh{}, \Mh{} and \Wh{}, with \Wh{} having an excess of perpendicular alignments, while \Lh{} and \Mh{} having the opposite, more aligned orientations. The fact that two thirds of the total population (\Mh{} and \Lh{} haloes) have spins typically aligned with \ec{} explains why this is the trend usually associated to low mass haloes. 

In Figure \ref{fig:gp_median_alignment_3x1} we show the median of $|\Jdir\cdot\ei|$, with $i=1,2,3$, as a function of halo mass. The solid lines correspond to the alignment at $z=0$. It shows that, at present time and for all masses, haloes show a clear correlation between their net spin growth and the spin--filament alignment. This correlation is largest for the alignment with filament axes \ea{} and \ec{}, and much smaller for \eb{}. For example, $|\Jdir\cdot\ec|$ is smallest for the \Wh{} sample (blue curve) indicating that those haloes have spins that are systematically more perpendicular on the filament spine, \ec{}, than the full halo population (black line) at all masses. Furthermore, the difference in $|\Jdir\cdot\ec|$ between the \Lh{} and \Wh{} samples is roughly the same at all halo masses and roughly equal with the variation in spin--filament alignment between the lowest and highest mass haloes shown in the figure. \textit{This indicates that net spin growth is as important as halo mass in determining how halo spins orient with respect to their host filaments.}

A commonly studied feature of the \Jdir{} alignment with \ec{} is the transition as a function of halo mass between an excess of parallel to an excess of perpendicular orientations, which is referred to as the spin transition mass. The bottom panel in Figure \ref{fig:gp_median_alignment_3x1} shows that this transition mass varies considerably as a function of net angular momentum growth and, in some case, no such mass can be defined. For example, the \Lh{} sample has an excess of parallel orientations for $\Mh\lesssim 10^{13}\hMsun$, and shows no preferential alignment at higher masses. Potentially, a transition towards perpendicular alignment could take place at higher masses, however we cannot probe this regime since most massive haloes are found in the nodes of the cosmic web and thus we rapidly run out of massive filament haloes. In contrast, the \Wh{} sample has an excess of perpendicular spin--\ec{} orientations at all masses and, if a spin transition is present, a linear extrapolation of our results suggest a value ${\sim}10$ times lower than the lowest mass halo resolved in our simulation.

Comparing the top and bottom panel in Figure \ref{fig:gp_median_alignment_3x1}, we notice that the spin alignment with \ea{} is reciprocal to that with \ec{}, that is the subsample which is more perpendicular to \ec{} is the one more parallel with \ea{}. This might seem obvious since \ec{} and \ea{} are, by definition, perpendicular to each other. However, it is absolutely non-trivial, because in principle the perpendicularity with respect to \ec{} could easily imply alignment with respect to \eb{}, or even a lack of alignment with either \ea{} or \eb. As we can clearly see in the central panel, the median alignment with respect to \eb{} not only has a marginal dependence on the mass, but also has a weak or negligible correlation with the haloes' spin growth, specially at high masses. Thus, the reciprocity between the median alignment with respect to the first and last axis of collapse of filaments suggests that the non-linear processes that affect the spin orientation have preferred directions within the $\ea-\ec$ plane.

It is instructive to compare the results in the bottom panel of Figure \ref{fig:gp_median_alignment_3x1} with the \citet{veenaetal2018,veenaetal2020} studies, which have shown that the halo spin--filament alignment depends on filament properties: at fixed mass, haloes have a higher tendency for \Jdir{} to be parallel to \ec{} if they reside in thick filaments \citep[see also][]{aragon2014}. We think that the two results, i.e. the median $|\Jdir\cdot\ec|$ trend with net spin growth and filament thickness, might be manifestations of the same effect. For example, it has been shown that \Lh{} haloes of low mass inhabit environments that are typically more clustered (see Figure 6 in \citealt{lopezetal2019}). Given that a significant fraction of low mass haloes is located away from the nodes of the cosmic web, this suggest that thick filaments are more commonly inhabited by \Lh{} haloes than by their \Wh{} counterparts. On the other hand, the difference in median $|\Jdir\cdot\ec|$ values between the \Lh{} and \Wh{} subsamples at the low mass end is ${\sim}0.13$ (0.56 versus 0.43), which is {\it 3 times} larger than the difference between median $|\Jdir\cdot\ec|$ in thick versus thin filaments (see Figure 14 in \citealt{veenaetal2018}). \textit{Thus, the halo net spin growth correlates more strongly with spin--filament alignments than the nature of filaments.}

\subsection{The $z=80$ spin--filament alignment for \Wh, \Mh, and \Lh{} haloes}
\label{sec:alignment_growth:spin_filament_IC}

\begin{figure*}
	\includegraphics[width=2\columnwidth]{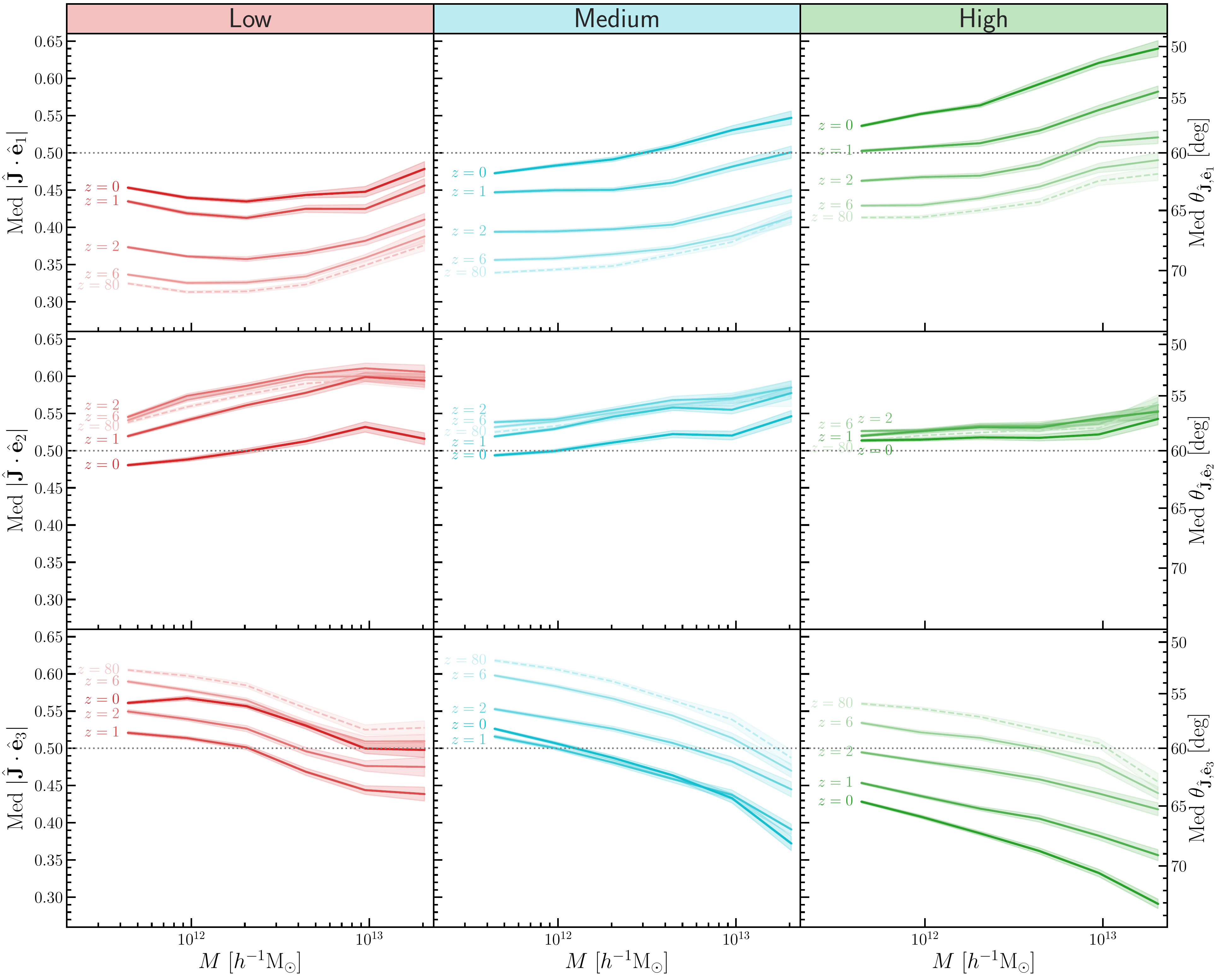}
    \caption{Same as in Figure \ref{fig:median_alignment_3x1}, but with the different columns corresponding to the samples derived from our classification. Each row shows the evolution of the median alignment between the angular momentum of \Lh{}, \Mh{} and \Wh{} haloes and one of the axis of their host filaments. The different curves in each panel represent the medians computed using the angular momentum directions at the given redshift and the axes directions fixed at present time. Shaded areas show the bootstrap $68\%$ confidence interval for each mass bin.}
    \label{fig:gp_median_alignment_3x3}
\end{figure*}

How is the present-day spin--filament configurations related to the TTT predictions? To address this question, we now study the alignment between the spin of proto-haloes' in the initial conditions and their $z=0$ filaments as a function of net spin growth. This is shown by the dashed curves in Figures \ref{fig:gp_alignment_dist_1x3} and \ref{fig:gp_median_alignment_3x1}. 

For example, the initial spin alignment with \ec{} (see right-most panel in Figure \ref{fig:gp_alignment_dist_1x3}) shows only a small dependence with net spin growth, at least when compared with the present day trend between \Jdir{} and \ec{} (see solid lines in the same figure). However, the same is not true for the alignment with \ea{}, where the initial and preset-day halo spins show the same degree of variation with halo net spin growth.

The same remarks are more easily visible in Figure \ref{fig:gp_median_alignment_3x1}, which shows the median alignment as a function of halo mass. For example, the $z=80$ spin alignment with \ea{} varies as a function of net spin growth to roughly the same extent as the present-day spin-\ea{} alignment. This is the case for all halo masses and it indicates that non-linear evolution, i.e. the change from $z=80$ to $z=0$, only shifts the overall median $|\Jdir\cdot\ea{}|$ value by the same amount for all haloes, independently of their net spin growth. In contrast, the spin alignment with \eb{} shows a more complex evolution: in the initial conditions there is a considerable trend with net spin growth that is reduced or even non-existent for massive haloes at present-day. In particular, at the low mass end, we find a reversal in the spin alignment trend with net spin growth: $|\Jdir\cdot\eb{}|$ at $z=80$ was highest for \Lh{} haloes while at $z=0$ is highest for \Wh{} haloes.

As stated above, the initial conditions spin alignment with \ec{} shows only a small variation with net spin growth and, in particular, the \Mh{} and \Lh{} samples have very similar median $|\Jdir\cdot\ec{}|$ values. The subsequent evolution, however, correlates strongly with the net spin growth, such that at $z=0$ we find considerable differences between the three samples (see section \ref{sec:alignment_growth:spin_filament_present-day}). There is a shift downwards (towards perpendicular configurations) that affects all samples, but that in principle seems to be more effective for haloes whose angular momentum grows the most. In effect, as we will see in the next section, the non-linear evolution of the spin filament alignment differs noticeably for \Lh{}, \Mh{} and \Wh{} haloes.

To summarize, we find a mixed picture of the drivers behind the dependence of the spin--filament alignment on net spin growth. In some cases, this difference is already present in the initial, i.e. TTT, spin orientation (e.g. alignment with \ea{}), while for other cases is due to the non-linear evolution of a halo's spin orientation being correlated with the net spin growth (e.g. alignment with \eb{} and \ec{}).

\subsection{Evolution of the spin--filament alignment for \Wh, \Mh, and \Lh{} haloes}
\label{sec:alignment_growth:spin_filament_evol}

Up to this point, we have found that the spin direction changes considerably from the initial conditions to the present-day and that some of these changes are strongly correlated with the net spin growth. In order to gain insight on how this evolution proceeds in time, Figure \ref{fig:gp_median_alignment_3x3} shows the median alignment between the spin and the preferred axes of the present-day filaments. Besides the $z=80$ and $z=0$ curves, which have already been shown in Figure \ref{fig:gp_median_alignment_3x1}, we show three intermediate redshifts, $z=6, 2$ and 1. Each column corresponds to one of the \Lh, \Mh{} and \Wh{} samples. 

We find that the spin alignment with \ea{} (top row in Figure \ref{fig:gp_median_alignment_3x3}) evolves at the same pace for the \Lh, \Mh, and \Wh{} samples, with only small differences. Hence, the fact that $|\Jdir\cdot\ea{}|$ varies with net spin growth at the present time is mostly due to the subsamples starting with different initial spin orientations with respect to \ea{}.

The alignment with \eb{} shows a more intricate evolution: the \Lh{}, \Mh{}, and \Wh{} samples start from slightly different initial values and they evolve at a different pace at late-times. The high redshift evolution is almost nonexistent and is the same for all subsamples until $z\sim2$, after which we find that subsamples peel off. For $z<2$, \Lh{} haloes show a slight but systematic decrease in $|\Jdir\cdot\eb{}|$, and this is also the case, to a minor extent, of \Mh{} haloes. The \Wh{} sample, on the other hand, shows a remarkably steady median alignment, which almost does not change neither at high or low redshift. 

Lastly, the evolution of the halo spin alignment with \ec{} is even more complex (see bottom row in Figure \ref{fig:gp_median_alignment_3x3}). All three subsamples decrease their $|\Jdir\cdot\ec{}|$ values, i.e. their spin becomes more perpendicular to \ec{}, by the same amount until $z\sim 2$. However, at very low redshift the change in $|\Jdir\cdot\ec{}|$ for \Lh{} haloes slows down and even reverses, such that at present time the \Lh{} subsample has the same alignment as at $z\sim6$. A similar trend can be seen for \Mh{} haloes, but the time of the reversal seems to take place later, such that their $z=0$ alignment is the same as the $z=1$ one. While not shown, we have checked that, between $z=1$ and $z=0$, the median $|\Jdir\cdot\ec{}|$ is not constant for the \Mh{} sample, but actually first keeps decreasing and then increases. The \Wh{} haloes are the only ones for which $|\Jdir\cdot\ec{}|$ systematically decreases at all redshifts, i.e. change towards more perpendicular orientations. 

The lack of change in the spin alignment with \eb{} for $z>2$, together with the changes observed in the alignment with \ea{} and \ec{}, indicate that in the linear and quasi-linear regime halo spin reorients itself by rotating along \eb{}, i.e. the intermediate axis of collapse on the scales relevant for filament formation. This process leads to spins more aligned with \ea{} and, thus, more perpendicular to \ec{}. Interestingly, \emph{this systematic and general evolution of the spin direction is not captured by TTT or other simple extensions of it.}
%-----------------------------------------------------------------------------------

\section{Summary and conclusions}
\label{conclusions}

%The last numbered section should briefly summarise what has been done, and describe the final conclusions which the authors draw from their work.

In this paper we have studied how the orientation of the spin and shape of DM haloes evolves with respect to the preferred axes of the filaments in the cosmic web, from the initial conditions to the present day. Our aim has been to better quantify one of the most important manifestation of environmental influence on halo and galaxy formation: the relation between the angular momentum of virialised haloes and the large-scale matter distribution of the Universe. The complexity of the halo spin--filament alignment has been demonstrated in previous studies \citep[e.g. see][]{leeypen2000,porcianietal2002a,porcianietal2002b,pazetal2006,aragoncalvoetal2007,hahnetal2007,codisetal2012,laigleetal2015,veenaetal2018,veenaetal2019,veenaetal2020}, but it remains unclear whether some of these results can be reconciled and interpreted from the perspective of the currently most accepted model for spin acquisition and evolution, the tidal torque theory (TTT).

To this end, we have used an N-body simulation with a large volume and high mass resolution. This has allowed us both to identify a statistically significant number of well resolved DM haloes at the present time, with masses between $\sim 10^{11}$ and $\sim 10^{14}\hMsun$ and to study their relation with the large-scale cosmic web. In order to analyze the evolution of this relation, we have followed back in time the particles of the present-day haloes up to their initial Lagrangian coordinates. In this process, we have calculated at different redshifts the properties of the corresponding proto-haloes', such as shape and spin orientation.

The identification of the filamentary network has been performed using the \nexus{} code \citep{aragon2007MMF,cautunetal2013,cautunetal2014}, which gives an unbiased and complete sample of filaments at different scales. In order to disentangle the evolution of proto-halo properties from changes in the filamentary network, we have analyzed the proto-haloes only with respect to the $z=0$ filamentary network. The latter, we have characterized in terms of its preferred axes, \ea{}, \eb{}, and \ec{}, which correspond to the directions of the first, secondary, and last large-scale collapse, respectively. In section \ref{sec:tidal_field_vs_filaments} we showed that the preferential axes of the initial tidal field, \ti{} with $i=1,2,3$, are well aligned with the axes of the present-day filaments.

From this analysis we obtained that:
\begin{itemize}
    \item The principal axes of the haloes' shape undergo a major change in their orientation with respect to the spine of the filaments, \ec{}. For instance, the major axis is strongly perpendicular at the initial conditions, but ends up being preferentially aligned at the present time, with high mass haloes slightly more aligned than their lower mass counterparts (see Figure \ref{fig:median_alignment_itF}).
    \item The halo spin--filament alignment changes significantly, even during the linear and quasi-linear regimes (e.g. $z>2$), in contrast to the TTT predictions that are often assumed to hold at such high redshift. 
    \item On median behaviour, the spin orientation of haloes evolves to become more aligned with \ea{} and more perpendicular to \ec{}, while there is hardly any change with respect to \eb{}, especially at $z>1$ (see Figure \ref{fig:median_alignment_3x1}).
\end{itemize}

One of our main goals has been to characterize how changes in the halo spin--filament alignment are related to the TTT. To this end, we have classified haloes according to their net spin growth, following the methodology introduced by \citet{lopezetal2019}. This led us to defining three subsamples: \Lh{}, \Mh{} and \Wh{} haloes, that correspond to objects whose angular momentum has grown, respectively, below, in consistency, or above the median expectation for their mass. This classification is correlated to the formation time of haloes, with \Lh{} haloes collapsing on average faster than the \Wh{} ones \citep{lopezetal2019}.

Using the \Lh{}, \Mh{}, and \Wh{} classification, we have reexamined the alignment between the spin and the preferred axes of the filaments to obtain the following results:
\begin{itemize}
    \item At the present time, our samples present remarkably different spin orientations with respect to the filament spine, \ec{}: \Lh{} haloes show a small excess of parallel orientations, while \Wh{} show the opposite, a clear excess of perpendicular spin--filament configurations (see Figure \ref{fig:gp_alignment_dist_1x3}).
    \item The difference in spin--filament alignment between the \Lh{} and \Wh{} samples at fixed halo mass is roughly equal to the change in alignment between the highest and lowest halo masses in our study. This highlights the crucial role of halo net spin growth in predicting how halo spins orient with respect to the cosmic web (see Figure \ref{fig:gp_median_alignment_3x1}).
    \item We find differences in the spin--filament alignment as a function of net spin growth (i.e. \Lh{} versus \Wh{} subsamples) already in the initial conditions. Non-linear evolution has a mixed effect on the differences between \Lh{} and \Wh{} haloes: mostly erases the ones in the spin alignment with \eb{} but enhances the differences in the spin alignment with \ec{}.
    \item At high redshift, $z>2$, the spin--filament alignment evolves at the same rate for \Lh{}, \Mh{} and \Wh{} samples towards more perpendicular configurations between \Jdir{} and \ec{}. At late times, \Lh{} and \Mh{} haloes peel off from this trend while \Wh{} haloes continue with it up to the present time (see Figure \ref{fig:gp_median_alignment_3x3}).
\end{itemize}

Our study reveals a prominent correlation between the halo spin growth and the spin orientation with respect to the preferential directions of the cosmic web. This adds to a growing body of works that reveal the complex processes shaping the relation between halo and galaxy rotation and the large-scale tidal forces generating them. For example, the spin--filament alignment depends on the scale on which filaments are identified \citep{codisetal2012,aragon2014,foreroromeroetal2014,Wang2018a} and, more importantly, on filament properties such as thickness and density \citep{veenaetal2018,veenaetal2020}. These trends highlight the intricate connection between spin acquisition and the tidal field at different spatial scales.

Our results show that net spin growth is an important indicator of halo spin orientation on par with halo mass and more significant than other secondary correlations. For example, the difference in median spin--filament alignment between our \Lh{} and \Wh{} samples is 3 times larger than the corresponding difference measured between thin and thick filaments \citep[compare with Figure 14 of][]{veenaetal2018}. This, in turn, suggests a strong correlation between spin--filament alignment and the collapse time of haloes, which is earliest for the \Lh{} sample \citep{lopezetal2019}.

The classification of haloes according to their net spin growth is intended to capture systematic deviations from the TTT predictions and thus could be argued that differences in the spin orientation of the various subsamples would arise only at late times. However, we find considerable differences in the spin orientations between \Lh{} and \Wh{} haloes already in the initial conditions. This is potentially due to the fact that the shape of a halo's Lagrangian volume, i.e. inertia tensor, is highly correlated with the local linear tidal field and also with the late-time non-linear tidal field that determines the collapse of a halo at low redshift \citep{vandeWeygaert1996,Ludlow2011,Rossi2013,Ludlow2014,Yu2020}. For example, \Wh{} haloes are more likely to correspond to somewhat spherical Lagrangian patches around density peaks in the initial conditions. In contrast, \Lh{} haloes are more likely to have formed in highly compressive tidal field regions and have more elongated initial shapes \citep{Borzyszkowski2017,lopezetal2019}.

Our study has revealed also changes in the halo spin orientation at high redshift, i.e. $z>2$, in mild disagreement with most TTT implementations, which predict constant spin directions. At early times, we find a systematic rotation of halo spins around the intermediate filament axis, \eb{}, such that the spin orientation becomes more aligned with the first filament axis, \ea{}. This rotation is the same for the \Lh{}, \Mh{}, and \Wh{} samples indicating a universal behaviour in the quasi-linear regime that is not captured by TTT. While the tidal field evolves slowly with time (see section \ref{sec:tidal_field_vs_filaments}), the same is not the case for the proto-halo shape, which undergoes a significant reorientation even in the mildly non-linear regime (see Figure \ref{fig:median_alignment_itF}). Thus, perhaps an extension of TTT that accounts for the evolution in the proto-halo shape could account for this high-redshift halo spin reorientation.

Such an extension is expected to fail first for \Lh{} haloes at $z\sim1$, which show a clear change in spin rotation before and after this redshift. For example, for $z>1$ the spin of \Lh{} evolves towards an excess of perpendicular orientations to \ec{}, and at later times this trend reverses completely to erase, or at least counteract, the angular momentum change produced by tidal torques during the linear and quasi-linear regimes. A similar behaviour is seen for the \Mh{} haloes too, but the reversal takes place at a lower redshift than for the \Lh{} sample. In contrast, no such behaviour is seen for the \Wh{} sample, potentially because these haloes collapse last. The late-time change in spin orientation seen for the \Lh{} and \Mh{} haloes could be due to late anisotropic inflow \citep{vanhaarlem&vandeweygaert1993,libeskindetal2013,codisetal2015,laigleetal2015,wangykang2017} or due to the emergence of strong vortical flows \citep{libeskindetal2012} with which haloes couple as they collapse and move from one environment to another. Besides modifying the spin direction, the net effect of such processes would be to reduce the total amount of angular momentum gained by these haloes and thus would lead to a clear correlation with our net spin growth classification.

One noticeable trend caused by this systematic behaviour is that, at the present time, one third of the total sample of low mass haloes have spins that are preferentially perpendicular to the spine of the filaments. More interestingly, these haloes belong to the \Wh{} sample. This does not necessarily contradict the widely reported preference of low-mass haloes to rotate with their axis aligned with the filament, but it is striking that there exists such a clear correlation between the present-day orientation of the spin and its growth in magnitude. Furthermore, \Wh{} haloes not only have the highest net spin growth for their mass, but they also are more rotationally supported than their \Lh{} and \Mh{} counterparts and show a remarkable coherence between their shape and spin direction \citep{lopezetal2019}. From a qualitative interpretation of a TTT-based model of spin acquisition such as ATTT \citep{codisetal2015}, one might naively expect that low mass haloes with these properties would have spins aligned with the filaments, since they would have formed from Lagrangian regions that occupy a single and coherent octant of vorticity. The fact that this is not the case constitutes one of the puzzling aspects of our results, and we would like to address this in a future work. An alternative interpretation could be that what ATTT adequately describes is, in fact, the mass trend of the spin-\ec{} alignment, which is already present in the initial conditions and remains almost unaltered up to the present time. In any case, a mechanism that fully explains the general shift towards more perpendicular configurations that we observe during the non-linear evolution remains to be found, specially if we want to understand certain aspects of the present day spin--filament alignment, such as the \emph{transition mass}.

\section*{Acknowledgements}
This work has been partially supported by the Argentine \emph{Agencia Nacional de Promoci\'on Cient\'ifica y Tecnol\'ogica} (ANPCyT), by means of the grant under reference PICT 2015-03098. Also, this project has received financial support from the European Union's Horizon 2020 Research and Innovation programme, under the Marie Sklodowska-Curie grant agreement No 734374: Project LACEGAL. PL would like to thank the hospitality of the Leiden Observatory (STRW) at Leiden University (NL), where part of this work has been done. 
MC acknowledges support by the EU Horizon 2020 research and innovation programme under a Marie Sk{\l}odowska-Curie grant agreement 794474 (DancingGalaxies). Numerical calculations were performed on the computer clusters at the \emph{Centro de C\'omputo de Alto Desempe\~no} (CCAD, http://ccad.unc.edu.ar) from the \emph{Universidad Nacional de C\'ordoba} (UNC). 

\section*{Data Availability}
The data used in this work is available upon reasonable request to the corresponding author.

%The Acknowledgements section is not numbered. Here you can thank helpful colleagues, acknowledge funding agencies, telescopes and facilities used etc. Try to keep it short.

%%%%%%%%%%%%%%%%%%%%%%%%%%%%%%%%%%%%%%%%%%%%%%%%%%



\bibliographystyle{mnras}
\bibliography{bibliografia} % if your bibtex file is called example.bib

\bsp	% typesetting comment
\label{lastpage}
\end{document}